\documentclass[fleqn,usenatbib]{mnras}

\usepackage[T1]{fontenc}

\DeclareRobustCommand{\VAN}[3]{#2}
\let\VANthebibliography\thebibliography
\def\thebibliography{\DeclareRobustCommand{\VAN}[3]{##3}\VANthebibliography}

\usepackage{graphicx}	
\usepackage{amsmath}	
\usepackage{amssymb}	
\usepackage{caption}
\usepackage{subcaption}

\usepackage{mwe}
\usepackage{newtxtext,newtxmath}

\title[GRB 201216C]{Jet-Cocoon Geometry in the Optically Dark, Very High Energy Gamma-ray Burst 201216C}

\author[L. Rhodes et al.]{
L. Rhodes$^{1, 2}$,\thanks{E-mail: lauren.rhodes@physics.ox.ac.uk}
A. J. van der Horst$^{3, 4}$,
R. Fender$^{1, 5}$,
D. R. Aguilera-Dena$^{6}$
J. S. Bright$^{7}$,
S. Vergani$^{8}$,
\newauthor
D. R. A. Williams$^{9}$
\\
$^{1}$ Astrophysics, Department of Physics, University of Oxford, Keble Road, Oxford OX1 3RH, UK\\
$^{2}$ Max-Planck-Institut f\"{u}r Radioastronomie, Auf dem H\"{u}gel 69, 53121 Bonn, Germany\\
$^{3}$ Department of Physics, the George Washington University, 725 21st Street NW, Washington, DC 20052, USA\\
$^{4}$ Astronomy, Physics and Statistics Institute of Sciences (APSIS), 725 21st Street NW, Washington, DC 20052, USA\\
$^{5}$ Department of Astronomy, University of Cape Town, Private Bag X3, Rondebosch 7701, South Africa\\
$^{6}$ Institute of Astrophysics, FORTH, Dept. of Physics, University of Crete, Voutes, University Campus, GR-71003 Heraklion, Greece\\
$^{7}$ Astronomy Department, University of California, Berkeley, CA 94720, USA\\
$^{8}$ GEPI, Observatoire de Paris, PSL University, CNRS, 5 Place Jules Janssen, 92190 Meudon, France\\
$^{9}$ Jodrell Bank Centre for Astrophysics, School of Physics and Astronomy, The University of Manchester, Manchester, M13 9PL, UK
}

\date{Accepted XXX. Received YYY; in original form ZZZ}

\pubyear{2022}

\begin{document}
\label{firstpage}
\pagerange{\pageref{firstpage}--\pageref{lastpage}}
\maketitle

\begin{abstract}
We present the results of a radio observing campaign on GRB 201216C, combined with publicly available optical and X-ray data. The detection of very high energy (VHE, >100\,GeV) emission by MAGIC makes this the fifth VHE GRB at time of publication. Comparison between the optical and X-ray light curves show that GRB 201216C is a dark GRB, i.e. the optical emission is significantly absorbed and is fainter than expected from the X-ray detections. Our \textit{e}-MERLIN data also shows evidence of diffractive interstellar scintillation. We can study the column density along the line-of-sight to the GRB in both the host galaxy, from the damped optical light curve, and the Milky Way, via scintillation studies. We find that the afterglow is best modelled using a jet-cocoon geometry within a stellar wind environment. Fitting the data with a multi-component model we estimate that the optical, X-ray and higher-frequency radio data before $\sim$25\,days originates from an ultra-relativistic jet with an isotropic equivalent kinetic energy of (0.6-10)$\times$10\textsuperscript{52}\,erg and an opening angle of $\sim$1-9$^{\circ}$. The lower-frequency radio emission detected by MeerKAT, from day 28 onwards, is produced by the cocoon with a kinetic energy that is between two and seven orders of magnitude lower (0.02-50)$\times$10\textsuperscript{48}\,erg. The energies of the two components are comparable to those derived in simulations of such scenarios.
\end{abstract}




\section{Introduction}
\label{sec:intro}

Long Gamma-ray Bursts (lGRBs) are flashes of gamma radiation usually lasting upwards of two seconds \citep{1993ApJ...413L.101K}. At cosmological distances, they are among the most energetic transients known. lGRBs are thought to be produced via internal processes in jets launched during the core collapse of a sub-population of fast rotating massive stars \citep[see ][for a recent review]{2016SSRv..202...33L}. Evidence for this connection has come from the presence of Type Ic broadline supernovae signatures in the optical afterglow spectra of lGRBs \citep[e.g.][]{2003Natur.423..847H}. 

Following the lGRB prompt emission, there is a broadband afterglow, often visible from radio frequencies to X-ray energies. The afterglow is produced when the jet interacts with the circumburst medium, creating two shocks: a forward and a reverse shock. The forward shock propagates into the surrounding medium whereas the reverse shock travels back into the jet and towards the newly formed compact object. The forward shock accelerates the electrons in the circumburst medium into a power-law distribution in energy: $N(E)dE \propto E^{-p}dE$, which subsequently cools emitting synchrotron and inverse Compton radiation. The synchrotron emission is interpreted in terms of the fireball model, the standard model used for GRB afterglows \citep{1992MNRAS.258P..41R}.

The synchrotron spectrum is described using four parameters: three break frequencies \citep{Sari1998} i.e. the self-absorption frequency ($\nu_{\textrm{SA}}$), the frequency corresponding to the electrons with the minimum energy ($\nu_{\textrm{m}}$) and the cooling frequency ($\nu_c$), and F\textsubscript{$\nu$, max}: the flux density at the peak of the spectrum, the higher of $\nu_{\textrm{m}}$ or $\nu_{\textrm{SA}}$. The three break frequencies are connected by power laws of flux as a function of frequency. The aforementioned four parameters evolve with time and they are dependent on the circumburst environment, its density and radial profile ( \textit{n} or \textit{A\textsubscript{*}}), the isotropic equivalent kinetic energy of the jet (E\textsubscript{K, ISO}), and the jet microphysical parameters, i.e. the fraction of the shock energy given to the electrons and magnetic fields \citep[$\epsilon_{e}$ and $\epsilon_{B}$, respectively ][]{1999ApJ...523..177W, 1999ApJ...520L..29C,2002ApJ...568..820G}. 

The reverse shock usually dominates between radio to optical wavelengths at early times, often decaying quickly within hours and days in the optical and radio wavebands, respectively \citep[e.g.][]{1999ApJ...522L..97K, 2012ApJ...746..156C, 2016ApJ...833...88L, 2016ApJ...833..100H, 2017ApJ...848...69A}. The forward shock emission is visible at X-ray energies from minutes to a few hours post burst \citep[e.g.][]{2009ApJ...698...43R, 2011MNRAS.412..561O}. After the reverse shock fades, the forward shock becomes visible at lower frequencies and can be visible for up to hundreds of days \citep{2008A&A...480...35V}.

As the peak of the synchrotron spectrum moves to lower frequencies with time, the afterglow light curves evolve chromatically as the frequency breaks pass through different observing bands at different times. At later times, some light curves also show achromatic breaks, caused by changes due to the jet's geometry and relativistic beaming of the emission. This is called a jet break and occurs in the regime where $\Gamma < 1/\theta_{j}$, where $\Gamma$ is bulk Lorentz factor of the of the forward shock and $\theta_{j}$ is the opening angle of the jet. The deceleration of the jet is such that the beaming cone widens allowing the observer to see the edges of the jet. As a result the observed flux of the jet begins to decay rapidly. Large sample studies of lGRB jet opening angles show jets to be highly collimated with $\theta_{j}$ =  7$^{\textrm{+11}}_{\textrm{-4}}$ $^{\circ}$ on average \citep{2014ApJ...781....1L}.

Observations of some GRB afterglows have shown evidence of a second forward shock component originating from a wider outflow \citep{2005A&A...440..477R,2005MNRAS.360..305S,2008Natur.455..183R,2009MNRAS.394..214K,2011A&A...526A.113F,2014MNRAS.444.3151V, 2018ApJ...860...44L,2020ApJ...891L..15C}. This second component is sometimes interpreted as a cocoon \citep[e.g.][]{ 2020ApJ...891L..15C}. Cocoons are often suggested to explain why GRB jets are so highly collimated. Magnetohydrodynamical numerical simulations show that as the relativistic jet propagates through the collapsing star, it deposits a large amount of energy into the surrounding material forming a cocoon. In turn, the cocoon reduces the lateral expansion of the jet resulting in a high degree of collimation when the jet breaks free \citep{2002MNRAS.337.1349R, 2004ApJ...608..365Z, 2018ApJ...863...32D}. The kinetic energy of the cocoon is expected to be several orders of magnitude less than the jet, more similar to relativistic supernovae, with a bulk Lorentz factor lower than the core of the jet \citep{2018ApJ...863...32D}. The signature of the cocoon can appear similar to that of the jet due to the cocoon's interaction with the circumburst medium producing synchrotron emission \citep{2017ApJ...834...28N}. However, the emission of this component is more likely observable in systems where the jet is viewed off-axis, or with a favourable combination of energetics and geometry, when the core jet cannot dominate the observed emission at all times. 

The afterglow is usually visible between the radio and X-ray wavebands, even reaching GeV energies in the most luminous events \citep{2013ApJS..209...11A}. However in the past three years, detections of the afterglow have been made at very high energies \citep[VHE, >100\,GeV][]{2019Natur.575..455M, 2021Sci...372.1081H}. There are now a handful of VHE detections made seconds to hours after the lGRB prompt emission. Their light curves are very similar to that observed in the X-ray band, implying a connection to the afterglow rather than the prompt emission \citep{2020GCN.28659....1B, 2019GCN.25566....1D}. Different production mechanisms have been invoked to explain the VHE emission: GRB 190114C has been best fit with a synchrotron self-Compton component \citep{2019Natur.575..455M}, whereas the GRB 190829A dataset has been best described using a single synchrotron emission component spanning from radio to VHE \citep{2021Sci...372.1081H}. 

VHE detections associated with lGRBs are limited in redshift due to pair production between the VHE photons and the extra-galactic background light: optical and infrared photons produced by star formation processes. As a result, VHE photons from sources above redshift 1.5 are highly attenuated, and are not expected to be detectable. This is supported by the VHE GRBs detected thus far: all at redshifts of about one or below \citep{2018GCN.22996....1V, 2019GCN.23708....1C, 2019GCN.25565....1V, 2020GCN.28661....1I}. We note that all of the VHE GRBs to date have been associated with strong radio detections \citep{2019Natur.575..455M, 2020MNRAS.496.3326R, 2020GCN.29028....1M, 2020GCN.28945....1R}.

Once the photons have left the GRB jet, they propagate through the inter-stellar medium (ISM) of the host galaxy, the inter-galactic medium and the ISM of the Milky Way. These media along the line of sight can dramatically affect the observed afterglow emission. Some lGRBs have appeared to be optically faint, so called dark GRBs. \citet{2012ApJ...746..156C} showed that about 25\% of Swift GRBs do not have detected optical counterparts. There are three possible explanations for dark GRBs \citep{2004ApJ...617L..21J, 2005A&A...440..477R}: (1) they occur at high redshift resulting in the Lyman break falling in the optical band, (2) dust along the line of sight absorbs the optical photons, and (3) additional emission components at X-ray energies, increasing the X-ray flux with respect to the optical emission. 

For many dark GRBs, it is dust along the line of sight that causes the optical darkness. Significant dust is expected in the regions of lGRBs as they occur in areas of high star formation, near to the birth-sites of their progenitor, since their lifetime is short. In some cases the extinction is greater than 10\,magnitudes \citep[V-band, e.g. ][]{2013ApJ...767..161Z}. When compared to the measured neutral hydrogen column densities inferred from the X-ray spectra, such high optical extinction deviates strongly from the linear A\textsubscript{V}-N\textsubscript{H} relationship measured within the Milky Way N\textsubscript{H}$\approx$ 2$\times$10\textsuperscript{21}A\textsubscript{V} \citep{1995A&A...293..889P, 2009MNRAS.400.2050G}. It is likely that this is a result of a combination of the A\textsubscript{V}-N\textsubscript{H} relation varying from galaxy to galaxy, in combination with a non-uniform distribution of gas within each galaxy. This is supported by observations of dark GRBs' host galaxies which appear to have `normal' colours, implying a lack of increased dust across the galaxy as a whole but rather localised to regions of increased star formation \citep{2009AJ....138.1690P}.

Material along the line of sight can also affect radio emission in the form of interstellar scintillation \citep[ISS;][]{1998MNRAS.294..307W, 1997NewA....2..449G}. Turbulence in the Milky Way's ISM causes flux density fluctuations at radio frequencies up to order unity. ISS can be divided into weak and strong scintillation. Weak scintillation occurs above some characteristic transition frequency. Strong scintillation occurs at frequencies below the transition frequency and can be further divided into diffractive (DISS) and refractive scintillation (RISS). DISS is a narrow-band effect resulting from multi-path propagation whereas RISS, a broadband effect, occurs due to the focusing and defocusing of rays as they propagate through the ISM. If observed, DISS will dominate at early times when the GRB jet is more compact, but as the size of the jet on the sky grows, the effects of DISS fade away leaving RISS \citep[][]{1997Natur.389..261F}. The effects of RISS will also quench at some time when the jet has expanded beyond a certain angular size on the sky. Depending on the observing frequency and angular size of the source, the variability can be up to 100\%. The angular size dependence of DISS and RISS can be used to place on constraints on the size of the jet at different epochs \citep{1997Natur.389..261F, 2000ApJ...534..559F, 2008ApJ...683..924C, 2019ApJ...870...67A}. 


\subsection{GRB 201216C}
\label{subsec:GRB201216C}
The prompt emission from GRB 201216C was detected on 2020 December 16 at 23:07:31 UT by the Neil Gehrels \textit{Swift} Observatory (here after \textit{Swift}) Burst Alert Telescope \citep[BAT,][]{2020GCN.29061....1B}. Three optical observatories also reported detections of a counterpart from early-time observations. A team searching for the afterglow with the Very Large Telescope (VLT) detected a source within the BAT error region at 21.81$\pm$0.05 magnitudes (r'-band) 2.19 hours after the burst \citep{2020GCN.29066....1I}. The VLT also measured a very steep optical spectral index ($\nu^{-4.1\pm0.2}$) and placed GRB 201216C at redshift z = 1.1 \citep{2020GCN.29077....1V}. \citet{2020GCN.29070....1J} and \citet{2020GCN.29085....1S} confirmed the optical source as the the afterglow. A number of other observatories reported deep upper limits \citep{2020GCN.29071....1O, 2020GCN.29210....1B, 2021GCN.29674....1G}. There was no report of a detection of a supernova component. The reported optical detections and upper limits are shown in Figure \ref{fig:lc} as the squares and downwards facing grey triangles, respectively.

The \textit{Swift}-X-ray Telescope (XRT) started observing $\sim$50\,minutes post burst. The XRT unabsorbed fluxes were very high with respect to the optical counterpart. When combined with the steep optical spectral index, GRB 201216C was classified as a dark GRB \citep{2020GCN.29077....1V}. It is unlikely that such a steep optical spectral index is a result of galactic extinction as the reddening in the direction of the burst is E(B-V) = 0.05 \citep{2020GCN.29071....1O}.  We discuss this further in Section \ref{subsec:dark}.

The Major Atmospheric Gamma Imaging Cherenkov (MAGIC) telescope, observing between 50\,GeV and 50\,TeV, reported the detection of a significant VHE counterpart less than a minute after the initial burst detection \citep{2020GCN.29075....1B}, making GRB 201216C is the highest redshift VHE GRB to date. Upon the notification of a VHE detection from the MAGIC Collaboration, we began a multi-frequency radio campaign with a series of successful Director's discretionary time observations (DDTs) with \textit{e}-MERLIN, the Karl G. Jansky Very Large Array (VLA) and MeerKAT. We also applied for late-time Target of Opportunity (ToO) observations  with \textit{Swift}-XRT. All of these observations are further discussed in Section \ref{sec:Obs}. In Section \ref{sec:results}, we present the temporal and spectral evolution of the afterglow of GRB 201216C. In Section \ref{sec:discussion}, we discuss the possible interpretations of the data using different jet models and discuss the ISM in both the Milky Way and the GRB's host galaxy.

\section{Observations}
\label{sec:Obs}

Here, we present the multi-frequency observations obtained and the data reduction process. A list of observing dates, peak flux density measurements and uncertainties are given in Table \ref{tab:rad_obs}. Spectral indices are only calculated in epochs where we have high enough signal to noise ratios.

\subsection{\textit{e}-MERLIN}

GRB 201216C was observed by \textit{e}-MERLIN three times through successful DDT proposals at 5, 12 and 29 days post burst (PI: Rhodes, project codes: DD10010 and DD11001). We observed at 5\,GHz with a bandwidth of 512\,MHz. Each epoch consisted of 60 minutes on the flux calibrator (3C286) and 90 minutes on the bandpass calibrator (OQ208) followed by eight hours of interleaved target and phase calibrators cycles: six minutes on the target and two minutes on the phase calibrator (J0056+1625). 

We used the \textit{e}-MERLIN pipeline to reduce the observations\footnote{\url{https://github.com/e-merlin/e-MERLIN\_CASA\_pipeline}}\citep{2021ascl.soft09006M}. The pipeline performs flagging, delay and bandpass calibration, and calculates phase and frequency dependent amplitude gain corrections which are all applied to the target field, along with flux density scaling from the flux calibrator. The calibrated measurement set was imaged in \textsc{casa} (Version 5.3.0) using the \textsc{tclean} task \citep{2007ASPC..376..127M}. The uncertainties associated with the flux density measurements combine the statistical uncertainty and a 5\% calibration error.

\subsection{Karl G. Jansky Very Large Array}

Six VLA observations were obtained through a DDT proposal (PI: Rhodes, project ID: 20B-456). We spread the observations out between 12 and 53 days post burst. The observations were made at 10\,GHz with a bandwidth of 4\,GHz. For each epoch, we observed the target field for 10 minutes, book-ended with the phase calibrator (J0121+1149) and the primary calibrator (3C147). We reduced the observations using the VLA pipeline in \textsc{casa} \citep[Version 5.3.0, ][]{2018AAS...23134214K}. The pipeline performs flagging, creates a model of the flux calibration, and performs initial calibration including antenna position corrections. Delay, bandpass and gain corrections are derived and applied to the data after which further flagging is performed. Imaging was also performed in \textsc{casa}. The uncertainties on the flux densities were calculated by combining the statistical error and 5\% calibration uncertainty added in quadrature.

\subsection{MeerKAT}
We obtained four DDT observations with MeerKAT (PI: Rhodes, DDT-20210107-LR-01) at 22, 29, 40 and 54 days post burst. Each observation lasted 140 minutes, made up of a five minute scan of a primary calibrator (J0408$-$6545) preceded by a series of 20 minute scans of the target interleaved with two minute scans of the secondary calibrator (J1808+0134). The observations were made at a central frequency of 1.28\,GHz with a bandwidth of 856\,MHz, split into 4096 channels. 

The MeerKAT data were reduced using \texttt{OxKAT}, a set of python scripts used for semi-automatic processing \citep{2020ascl.soft09003H}. Firstly, the calibrator fields were flagged for RFI as well as the first and last 100 spectral channels. A spectral model from the primary calibrator was applied to the secondary. Delay, bandpass and complex gain calibration was performed on the primary and secondary calibrators and applied to the target field. Finally the target field was flagged using \textsc{tricolour}\footnote{https://github.com/ska-sa/tricolour}. The data were imaged with \textsc{wsclean} using a Briggs weighting with robust parameter of -0.7 \citep{2014MNRAS.444..606O}. We derived a model from the image and used it to reimage after a round of phase-only self calibration. The flux uncertainties include statistical uncertainties and a 10\,\% calibration error.

\begin{table}
    \centering
    \begin{tabular}{c c c c c}
    \hline
         T-T$_{\rm o}$ & $\Delta$T & $\Delta\nu$ & S$_{\nu}$ & $\beta$ \\
         (days) & (hours) & (GHz) & ($\rm \mu$Jy) & \\
         \hline
         \multicolumn{5}{c}{\textit{\textit{e}-MERLIN}} \\
         \hline
         5.6 & 8 & 4.8$-$5.2 & 180$\pm$23 & 7$\pm$4\\
         21.9 & 6 & 4.8$-$5.2 & $<$102 & - \\
         28.7 & 6 & 4.8$-$5.2 & 66$\pm$10 & - \\
         \hline
         \multicolumn{5}{c}{\textit{VLA}} \\
         \hline
         12.1 & 0.2 & 8$-$12 & 127$\pm$12 & 1.8$\pm$0.8 \\
         14.0 & 0.2 & 8$-$12 & 98$\pm$6 & 1.5$\pm$0.8 \\
         20.0 & 0.2 & 8$-$12 & 124$\pm$13 & 0.4$\pm$0.6\\
         36.0 & 0.2 & 8$-$12 & 62$\pm$6 & - \\
         44.0 & 0.2 & 8$-$12 & 68$\pm$9 & - \\
         53.0 & 0.2 & 8$-$12 & 50$\pm$5 & - \\
         \hline
         \multicolumn{5}{c}{\textit{MeerKAT}} \\
         \hline
         22.8 & 2 & 0.9$-$1.7 & $<$29 & -\\
         28.7 & 2 & 0.9$-$1.7 & 95$\pm$11 & $> -$0.3\\
         40.6 & 2 & 0.9$-$1.7 & 124$\pm$15 & $-$1.1$\pm$0.6\\
         54.5 & 2 & 0.9$-$1.7 & 130$\pm$14 & $-$0.7$\pm$0.5\\
         \hline
    \end{tabular}
    \caption{A table of the radio observations made with \textit{e}-MERLIN, the VLA and MeerKAT. The columns are the following: T-T\textsubscript{0}, the time between the burst detection and the
centre of the observation, in days; $\Delta$T, the duration of the observation, in hours; $\Delta\nu$, the observing frequency range; S$_{\nu}$, the peak flux density (or 3$\sigma$ upper limit); $\beta$, the in-band spectral index. The uncertainties on each flux density measurement are a combination of the fitting error
and a calibration error (5$\%$ for \textit{e}-MERLIN and VLA, 10$\%$ for MeerKAT) added in quadrature. We only give values for $\beta$ for epochs when the source is
bright enough to be detected in at least one half of the band.}
    \label{tab:rad_obs}
\end{table}

\subsection{Neil Gehrels Swift Observatory - X-ray Telescope}

The \textit{Swift} X-ray Telescope (XRT) observed the field of GRB 201216C from 3000\,s until 22\,days after the initial burst \citep{2021GCN.29280....1E}. This included two ToO observations that we obtained between days 20 and 27 post burst. Each epoch was automatically fitted with a power-law spectrum. The light curve and spectra are made public on the \textit{Swift} Burst Analyser \citep{Evans2007, Evans2009, Evans2010}. The X-ray flux densities used in our analysis are calculated at 5\,keV to avoid systematic under or over estimations in calculating the flux density at the edge of the observing band (i.e. at 0.3 or 10\,keV).

\section{Results}
\label{sec:results}

In the following sections, we use the convention $F_{\nu} \propto t^{\alpha}\nu^{\beta}$ where \textit{t} is the time post burst, $\nu$ is the observing frequency, and $\alpha$ and $\beta$ are the exponents. Any subscripts are used to indicate the relevant part of the spectrum or frequency band. 

\subsection{Light Curves}

\begin{figure*}
    \centering
    \includegraphics[width = \textwidth]{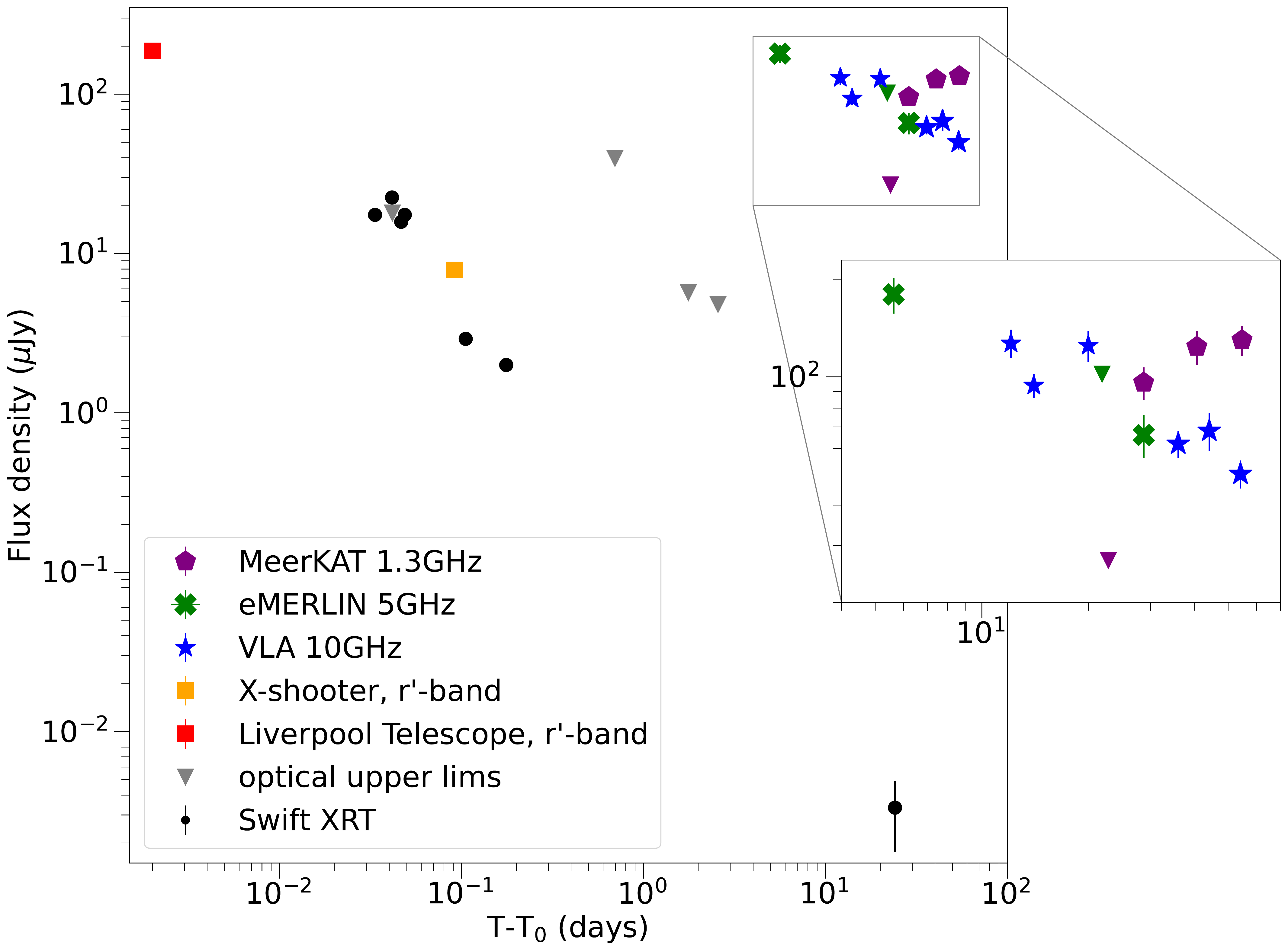}
    \caption{X-ray, optical and radio observations from GRB 201216C. The flux densities for the radio data points are given in Table \ref{tab:rad_obs}. The optical flux densities and upper limits are from the Gamma-ray burst Coordinates Network Circulars \citep{2020GCN.29085....1S, 2020GCN.29066....1I,2020GCN.29210....1B,2021GCN.29674....1G}. The \textit{Swift}-XRT light curve for GRB 201216C has been rebinned into five minute bins. The inset shows a clearer view of the radio dataset. }
    \label{fig:lc}
\end{figure*}

\subsubsection{Radio}
\label{subsubsec:radio}

Figure \ref{fig:lc} shows the radio light curves from our observing campaign. The flux densities and upper limits are given in Table \ref{tab:rad_obs}. We detected radio emission at 5\,GHz in two of the three observations with \textit{e}-MERLIN, during epoch one and three (the green crosses in Figure \ref{fig:lc}).

Our 10\,GHz light curve (blue stars in Figure \ref{fig:lc}) from the VLA covers the largest time range, from 12 to 54 days post burst. The 10\,GHz behaviour is best described as a shallow power-law decay ($\alpha_{\textrm{10\,GHz}} = -0.5\pm0.1$) from $\sim$120$\mu$Jy at 12 days to $\sim$50$\mu$Jy at 54 days. On top of the decaying flux, we detect inter-observation variability, which is possibly due to ISS (see Section \ref{sec:scint}). The epoch-to-epoch variability could cause the observed decay rate to deviate significantly from the intrinsic evolution. 

The 1.3\,GHz MeerKAT light curve (the purple stars in Figure \ref{fig:lc}) starts with a very steep rise ($\alpha_{\textrm{1.3\,GHz}}\gtrsim5$) from a 3$\sigma$ upper limit of 29$\mu$Jy to a detection of 95$\mu$Jy in 6\,days. The next three data points show a fairly flat light curve ($\alpha_{\textrm{1.3\,GHz}} = 0.1+/-0.2$) to 130$\mu$Jy in the final epoch. The sharpest rise possible for the standard forward shock model is $t^{1.75}$, which is still far shallower than the observed rise, comes from optically thick synchrotron from a forward shock propagating through a stellar wind environment in the regime where $\nu_{\textrm{m}}\textrm{ < }\nu_{obs}\textrm{ < }\nu_{\textrm{SA}}$ \citep{2002ApJ...568..820G}. 

\begin{figure*}
    \centering
    \includegraphics[width = 0.9\textwidth]{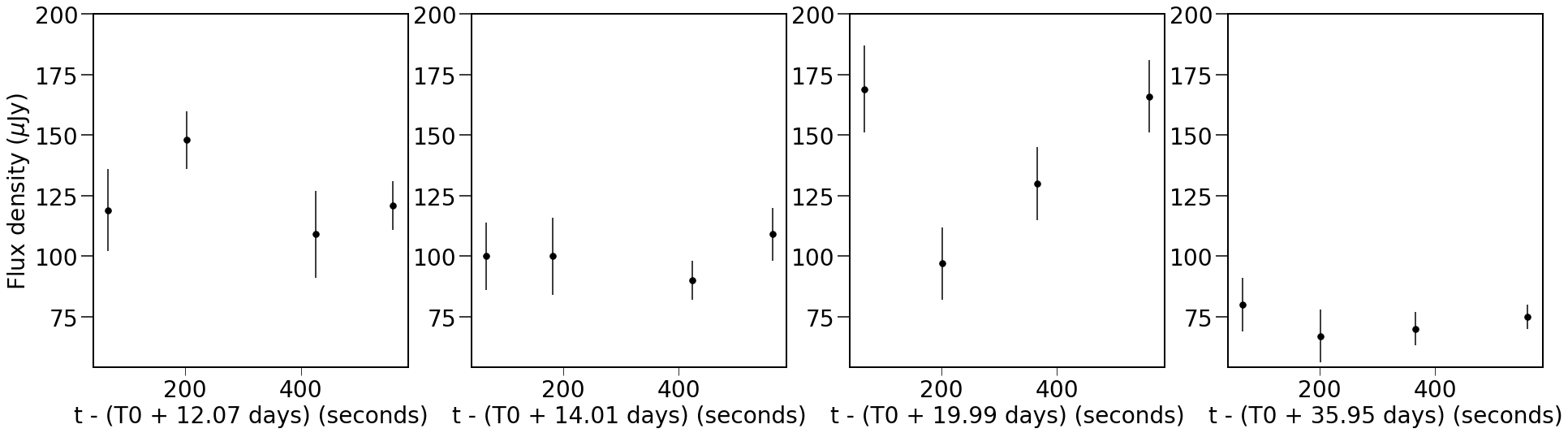}
    \caption{Intra-observation light curves to show short term variability in our VLA 10\,GHz data set. The first four 10 minute epochs are sub-divided into four-2.5 minute sub-integrations. We only use the first four observations as these are the brightest four where the source is reliably detected on short timescales.}
    \label{fig:vla_scint}
\end{figure*}

Due to the jet's compactness, radio observations of GRB afterglows are susceptible to scintillation, which can cause significant spectral and temporal variability, especially at early times. We note that scintillation timescales are also frequency dependent and that we are sampling variability on different timescales with the different interferometers due to differing observation lengths. ISS may be the cause of the inter-observation variability seen with the VLA. We also search for intra-observation variability within the VLA data, as shown in Figure \ref{fig:vla_scint}. We observe some low-level variability, $\sim$10-20\% \citep[using equation 10 from ][]{2003MNRAS.345.1271V}, in VLA epochs one and three. VLA epochs two and four show no such variability (see Figure \ref{fig:vla_scint}). The flux densities of the last two observations are too low to search for variability.

\begin{figure}
    \centering
    \includegraphics[width=\columnwidth]{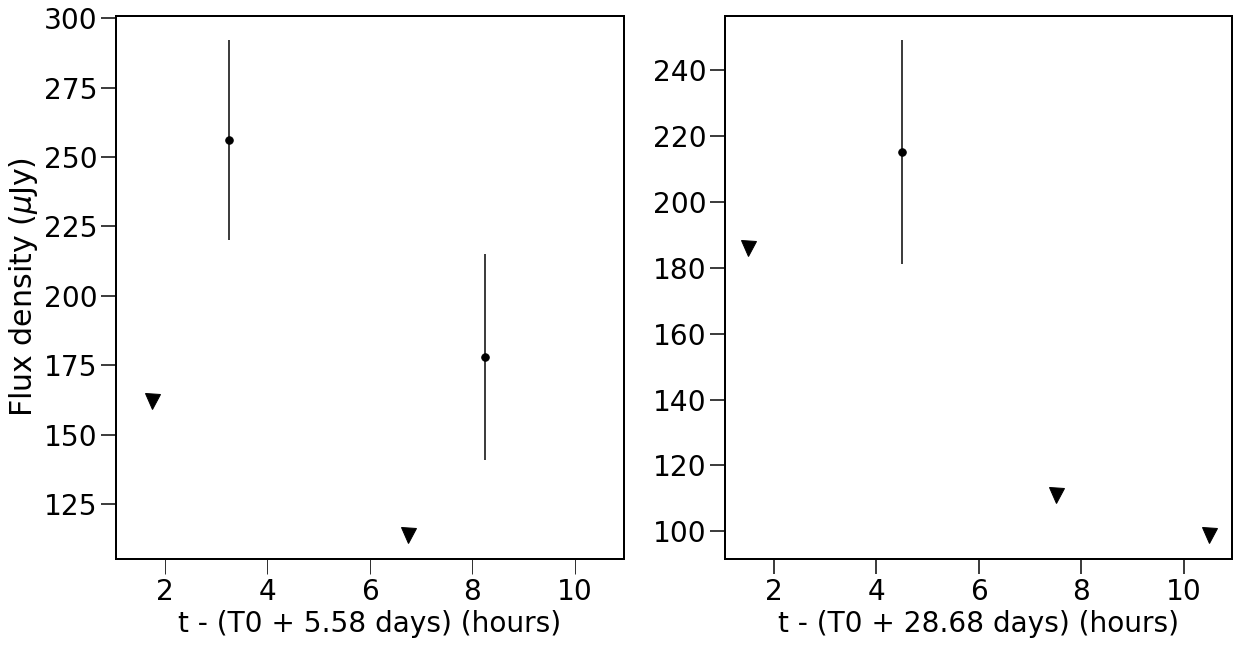}
    \caption{Short term variability observed in the first and third \textit{e}-MERLIN epochs. Each six hour observation was split into four sub-integrations. In the first observation made 6\,days post burst, we detected radio emission in two of the four sub-integrations. In the third observation, made at 28 post burst, we only detected the source in one of the four sub-integrations. }
    \label{fig:emerlin_scint}
\end{figure}

We split the two \textit{e}-MERLIN detections, which are six hours long each, into four-90 minute segments. The left hand panel of Figure \ref{fig:emerlin_scint} shows that for the first epoch, radio emission was only detected in two of the four segments. In the final epoch, we only detect radio emission for 90 minutes out of six hours.  

Our MeerKAT data set shows no evidence of intra-observation variability on a timescale of tens of minutes. The implications of the observed variability are discussed further in Section \ref{sec:scint}.

\subsubsection{Optical}
\label{subsubsec:opt_lc}

An optical counterpart to GRB 201216C was detected with the Liverpool telescope and the VLT \citep[red and orange squares in Figure \ref{fig:lc}, respectively; ][]{2020GCN.29085....1S, 2020GCN.29066....1I}. Within a couple of hours, the source had faded below detection limits. The r'-band light curve follows a $\alpha_{r} = -0.83\pm0.01$ decay from a few minutes post burst \citep{2020GCN.29210....1B,2021GCN.29674....1G}. Such a decay rate is consistent with optically thin synchrotron radiation in an ISM environment which gives p = 2.11$\pm$0.01 (from \citep[$\alpha = 3(1-p)/4$, ][]{2002ApJ...568..820G}. An optically thin forward shock in an wind environment would decay more rapidly (for p $=$ 2, $\alpha = -1.25$), however, with only a handful of detections it is impossible to determine whether the measured decay rate could be due to a combination of a frequency break passing through the optical observing band and a stellar-wind-like environment. Such a combination would result in the optical light curve appearing shallower.

\subsubsection{X-ray}
\label{subsubsec:X-raylc}

The XRT light curve (black circles in Figure \ref{fig:lc}) consists of a significant number of detections early on (t < 1\,day) and one further detection around 25\,days post burst. We have rebinned the photon-counting mode light curve to reduce any bias towards the earlier detections when fitting a power law decay to the light curve. Using either five or ten minute bins, we obtain a decay rate of $\alpha_{X-ray} = -1.9\pm0.3$. The final detection is slightly above the predicted flux density given the above decay rate as shown in Figure \ref{fig:forward_shock}.


There are many scenarios in which the X-ray light curve agrees with theoretical predictions. (1) Synchrotron radiation from a forward shock above $\nu_{\textrm{C}}$ in either a homogeneous or wind environment, $\alpha = (2-3p)/4$, giving p = 3.2$\pm$0.4; (2) optically thin forward shock synchrotron emission below $\nu_{C}$ in a stellar wind environment, $\alpha = (1-3p)/4$, giving p = 2.9$\pm$0.4 \citep{2002ApJ...568..820G}; (3) optically thin forward shock emission below $\nu_{C}$ in a homogeneous environment, $\alpha = 3(1-p)/4$: p = 3.5$\pm$0.4, and (4) the early X-ray light curve is also consistent with a jet break with no significant lateral spreading, with only edge effects considered. In a homogeneous environment, emission above the cooling break should decay as $\alpha = -(1+3p)/4$, where p = 2.2$\pm$0.2, and below the cooling break $\alpha = -3p/4$, where p = 2.5$\pm$0.2. In a stellar wind environment, emission below the cooling break should decay as $\alpha = -(1+3p)/4$, where p = 2.2$\pm$0.2, and above the cooling break $\alpha = -3p/4$, where p = 2.5$\pm$0.2 \citep{2013NewAR..57..141G}. We can only break the degeneracy between the different potential scenarios by considering the X-ray spectral index measurements and light curves in other wave bands to form a broadband model.

\subsection{Short timescale radio variability}
\label{sec:scint}

As well as the long term evolution, our observations also show evidence of short timescale (inter- and intra-epoch) variability as a result of ISS.

\subsubsection{Diffractive scintillation}
\label{subsubsection:eMERLINdiff}

DISS causes narrow-band fluctuations of order unity on a range of timescales and therefore, can affect radio observations dramatically. Our observations from \textit{e}-MERLIN show evidence of short timescale variability, a feature that is inconsistent with the smooth variations expected from GRB afterglows. Here, we explain that the variability observed is a result of small scale inhomogeneities in the local ISM which causes multi-path propagation (DISS) of the radio waves from GRB 201216C \citep{1997NewA....2..449G}. For simplicity, the region of the ISM causing the scattering is collapsed into a screen as some distance along the line of sight \citep{1998MNRAS.294..307W}.

We placed a lower limit on the intra-observation temporal variability of 30\% for our \textit{e}-MERLIN data set \citep{2003MNRAS.345.1271V}. If the observed variability is due to DISS, we would expect to see narrow-band flux modulations up to one on the timescale of an hour \citep{1997NewA....2..449G, 1998MNRAS.294..307W}. Due to signal to noise limitations, we cannot search for shorter timescale variability. We also searched for variability in the spectral domain by dividing the two detection epochs into 4 sub-bands (centred at 4.8, 4.9, 5.1 and 5.2\,GHz), another sign of DISS since DISS is a narrow-band phenomenon. In the first \textit{e}-MERLIN epoch, radio emission is only detected in the sub-band centred at 4.8\,GHz, at 300$\pm$30$\mu$Jy. In the bands centred at 4.9, 5.1 and 5.2\,GHz, we obtained 3$\sigma$ upper limits of 156, 144 and 195$\mu$Jy, respectively. A high level significance detection in only a single, narrow frequency band implies that DISS is most likely the origin of the variability at 5\,GHz. In the final \textit{e}-MERLIN epoch, the low flux density of the source means we are unable to detect emission in any of the sub-bands. 

Under the assumption that the \textit{e}-MERLIN variability is caused by DISS, we are able to place constraints on the location of the scattering screen between the Earth and the position of the GRB. If we assume that the screen is located within the Milky Way, by integrating the free electron distribution along the line of sight using the NE2001 model \citep{2002astro.ph..7156C}, we calculate a scattering measure SM\textsubscript{-3.5} $=$ 0.69 where SM\textsubscript{-3.5} = SM/($10^{-3.5}\,\textrm{kpc}~\textrm{m}^{-20/3}$), defined as the characteristic angle that incoming radio waves are scattered by whilst propagating through the ISM and from it infer a transition frequency of 8.8\,GHz. Observations below 8.8\,GHz are in the regime where it is possible to observe strong scattering, consistent with our conclusion that the variability at 5\,GHz is produced by DISS. This scattering measure and transition frequency correspond to a scattering screen at a distance ($d_{\textrm{scr}}$) of $\sim$0.9\,kpc. We obtain the same value for $d_{\textrm{scr}}$ using the method presented in \citet{1997NewA....2..449G}.

DISS is heavily dependent on the angular size of the GRB. Once the GRB jet has expanded above a critical size on the sky, the effects of DISS will no longer will observable. This critical size is determined by the distance to the scattering screen, observing frequency, and the scattering measure \citep{1997NewA....2..449G}:
$$\theta_{s} < 2.25 \nu_{10}^{6/5} (SM_{-3.5})^{-3/5} d_{\textrm{scr, kpc}}^{-1} \mu\textrm{as}$$

We know that the radio emission observed is still affected by DISS 29 days post burst, based on the short timescale variability observed in our last \textit{e}-MERLIN observation (Figure \ref{fig:emerlin_scint}). Therefore, at 29 days post burst the angular size of the jet associated with GRB 201216C must be less than 1$\mu\textrm{as}$. At a redshift of 1.1, the distance to GRB 201216C, 1$\mu\textrm{as}$ is $\sim 1\times10^{17}$cm. This size upper limit is consistent with size measurements of other lGRBs made at around 30 days \citep[see figure 7 of][]{2019ApJ...870...67A, 2004ApJ...609L...1T}. It is likely that if subsequent observations at 5\,GHz were made, they would most likely not have been affected by DISS. 

According to the thin screen scattering model, at 1.3\,GHz (the MeerKAT observing band), we would expect to see variability on timescales of 10 minutes with a modulation index of one. We see no intra-observation or narrow-band variability in our MeerKAT data. This indicates that by 29 days post burst, the angular size of the jet has grown larger than the 0.3$\mu\textrm{as}$ ($0.3\times10^{17}$cm).

\subsubsection{Refractive scintillation}

The presence of DISS at 5\,GHz also implies the presence of RISS. RISS produces variability on longer timescales. At 1.3 and 5\,GHz, we would expected to see variability at a level of about 30 and 70\% on timescales of over 5\,days and $\sim$7\,hours, respectively \citep{1998MNRAS.294..307W}. It is possible to use RISS to also constrain the source size:

$$\theta_{\textrm{S}} < 8 \nu_{0}^{17/10} \nu^{11/5} \textrm{d}_{\textrm{scr,kpc}}^{-1/2}$$

where $\nu_{0}$ and $\nu$ are the transition and observing frequency. respectively.

The observations at 5\,GHz are dominated by the effects of DISS, and due to the sparse cadence, days between each epoch, we are unable to observe the effects of RISS. In the MeerKAT band, the increase in flux density between days 23 and 29 is greater than a factor of three, far higher than the predicted RISS flux modulations of 30\%. The observations in which the source detected shows a smooth increase in flux density across the three epochs in which we detect the source. The spacing between each epoch is too large to infer whether the increase in flux density is due to RISS. Therefore, we cannot confidently attribute the MeerKAT flux variations to RISS.

\subsubsection{Weak Scintillation}

Weak scintillation often affects the data at a level similar to that of the calibration uncertainties ($\sim5-10\%$); although it can be significantly higher for observing frequencies close to the transition frequency). Our observations at 10\,GHz, above the transition frequency, show clear inter-observation variability (the blue stars in Figure \ref{fig:lc}), as well as at the $\sim$10\% level on the timescale of minutes in the two of the first four epochs, see Figure \ref{fig:vla_scint}. The flux density of the radio counterpart in the last two VLA epochs are too low to search for intra-observation variability. As a result, we cannot tell if the effects of weak scintillation have faded as the jet grows on the sky. 

With a transition frequency of 8.8\,GHz, the majority of the VLA observing band falls within the weak scattering regime. However, the VLA's wide bandwidth means we probably still observe some effects of DISS and RISS. DISS, RISS and weak ISS are expected to cause variation on timescales of two to three hours across the VLA band with a modulation index as high as 1 \citep{1998MNRAS.294..307W, 2014PASA...31....8G}. Such high variability levels are to be expected because our observations are so close to the transition frequency, although the amplitude of the flux modulation is expected to drop rapidly towards high frequencies. Therefore, it is most likely that the variability observed in the VLA band is a combination of DISS and weak scintillation. 

\subsection{Spectra}

\subsubsection{Radio}

The right-most column of Table \ref{tab:rad_obs} shows all the in-band radio spectral index measurements calculated using the individual observing bands for epochs where the source was bright enough. Of the three \textit{e}-MERLIN observations, only the first epoch was bright enough to obtain an in-band spectral index. The scintillation dramatically affects the \textit{e}-MERLIN spectra as it does the intra-epoch light curves. At 5 days post burst, the only time where the source is bright enough to split the band in two, we measure a $4.8-5.3$\,GHz spectral index of 7$\pm$4. The large uncertainties mean that such a steep result is still compatible with the steepest branch of the synchrotron spectrum in the GRB afterglow scenario, caused by synchrotron self-absorption. 

The radio emission at 10\,GHz (VLA) is only bright enough in the first three epochs to split the 4\,GHz bandwidth into two-2\,GHz subbands. In each of these three epochs, the 8-12\,GHz spectral index is consistent with being spectrally steep or fairly flat ($\beta_{\textrm{10\,GHz}}\geq$0). Over the course of the three observations, the VLA in-band spectral index slowly flattens (see Table \ref{tab:rad_obs}). The wide VLA observing band smears out any narrow-band effects of DISS.  In the context of the fireball model, the observations made at 12 and 14 days post burst are too steep to be in the regime where $\nu_{\textrm{SA}}$ < 10\,GHz < $\nu_{\textrm{m}}$ ($\beta = 1/3$). Instead, they are more consistent with 10\,GHz < $\nu_{\textrm{SA}}, \nu_{\textrm{m}}$ ($\beta = 2$) or $\nu_{\textrm{m}}$ < 10\,GHz < $\nu_{\textrm{SA}}$ \citep[$\beta = 2.5$,][]{2002ApJ...568..820G}.

We also calculated the 1.0-1.7\,GHz spectral index for the three MeerKAT observations where we detect radio emission (values are also given in Table \ref{tab:rad_obs}). At 28 days post burst, we were only able to detect radio emission in the upper half of the band, so we only have a lower limit on the spectral index, here the spectral index is approximately flat. The final two epochs show a spectral index of $\beta_{\textrm{1.3\,GHz}}$<0, consistent with optically thin synchrotron:  $\nu_{\textrm{m}}$, $\nu_{\textrm{SA}}$ < 1.3\,GHz. The three MeerKAT detections are made after the epochs where the 8-12\,GHz spectral indices are calculated, meaning that over the course of the radio campaign, the emission evolves from being optically thick to optically thin.

\subsubsection{X-ray}

The \textit{Swift}-XRT spectra in the range of 0.3-10\,keV are each fitted with a power law parameterised by the photon index: $\Gamma$, where $\beta_{X} = 1-\Gamma$. There are no significant variations in $\Gamma$ over the observing period implying that no break frequency passes through the XRT band. The data taken in photon counting mode result in $\Gamma = 2.0\pm0.1$ ($\beta_{X}$ = -1.0$\pm$0.1). As with the X-ray light curve, $\beta_{X}$, emission both below (p = 3.0$\pm$0.2) and above (p = 2.0$\pm$0.2) $\nu_{\textrm{C}}$. The spectral index does not change in the event of a jet break, unlike the light curves.

\subsubsection{Broadband spectra}
\label{sec:broadband}

    \begin{figure}
        \centering
        \begin{subfigure}[b]{0.45\textwidth}  
            \centering 
            \includegraphics[width=\textwidth]{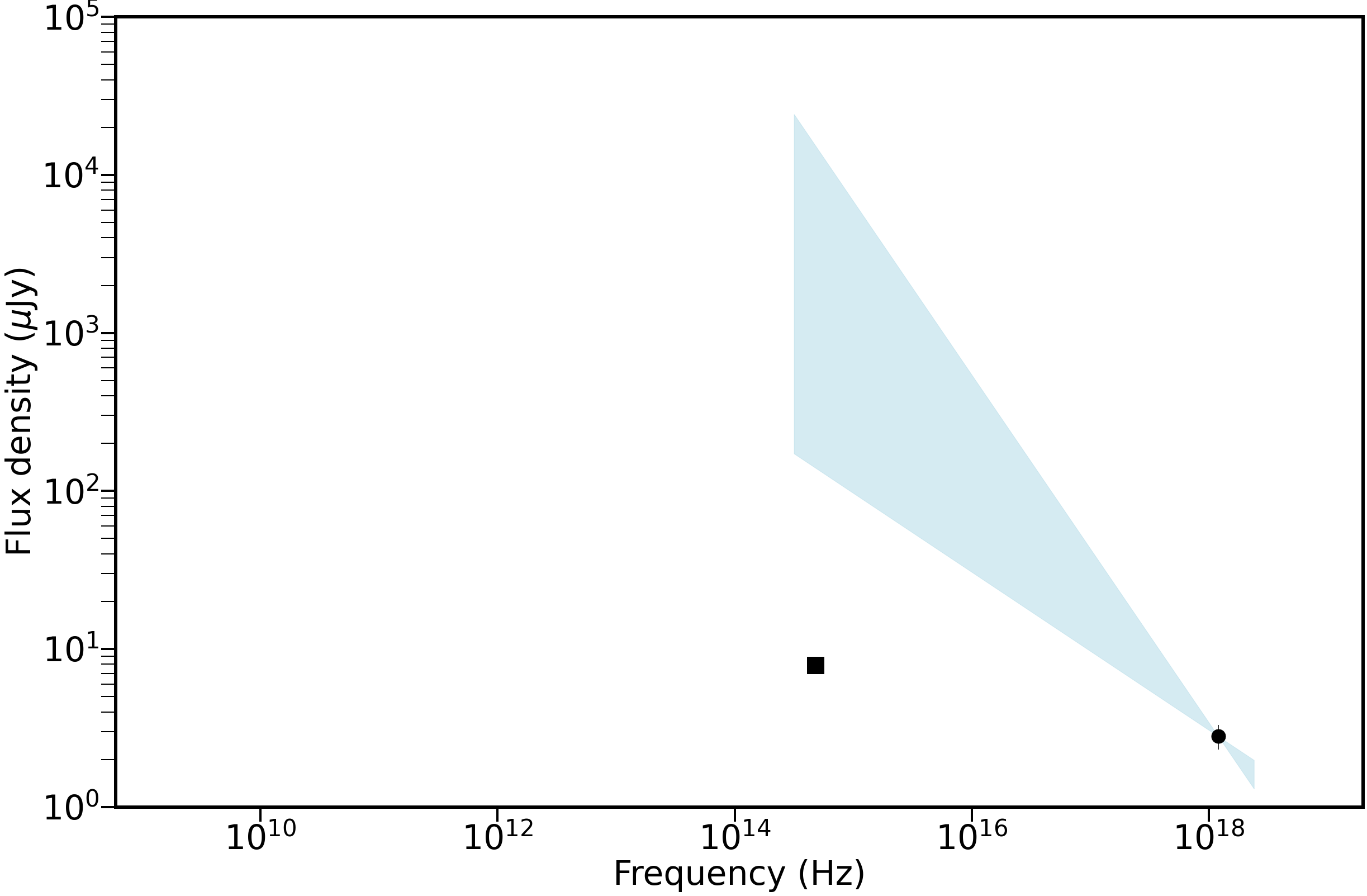}
            \caption[ ]%
            {{\small Optical and X-ray detections from 2 hrs post burst. The blue shaded region corresponds to the range in predicted spectra assuming that $\nu_{\textrm{C}}$ falls between the optical and X-ray bands. Comparison of the faint optical detection and the hypothetical synchrotron spectrum shows that GRB 201216C is a dark GRB. 
            \vspace{0.35cm}}} 
            \label{fig:mean and std of net24}
        \end{subfigure}
        \vspace{-0.2cm}
        \vskip\baselineskip
        \begin{subfigure}[b]{0.45\textwidth}   
            \centering 
            \includegraphics[width=\textwidth]{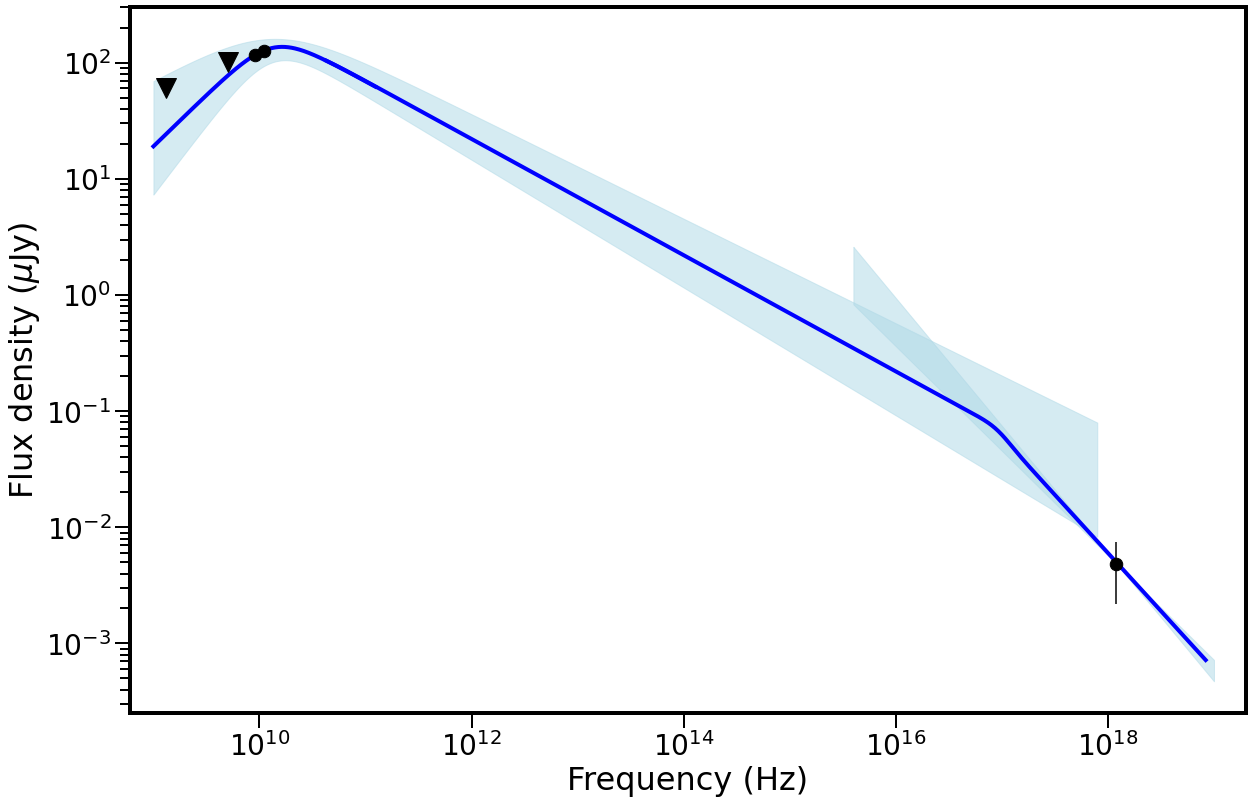}
            \caption[]%
            {{\small Broadband spectrum from 0.9\,GHz to 10\,keV using data from MeerKAT, \textit{e}-MERLIN, VLA and \textit{Swift}-XRT made between 20 and 24 days. We use the \textit{Swift}-XRT spectral index to calculate p, which is then applied to construct the spectrum above the peak frequency. We fit a power law to the radio data ($\beta_{\textrm{rad}} = 0.9\pm0.4$), assuming that the peak of the spectrum is at/above the VLA band. The shaded regions reflect the 1$\sigma$ uncertainties on the spectral indices. We infer that $\nu_{\textrm{C}}$ is between 8$\times$10\textsuperscript{15} and 8$\times$10\textsuperscript{17}\,Hz and that the peak of the spectrum at 13$\pm$9\,GHz. We measure F\textsubscript{$\nu$, max} to be 130$\pm$30\,$\mu$Jy.}}
            \label{fig:mean and std of net34}
        \end{subfigure}
        \vskip\baselineskip
        \begin{subfigure}[b]{0.45\textwidth}   
            \centering 
            \includegraphics[width=\textwidth]{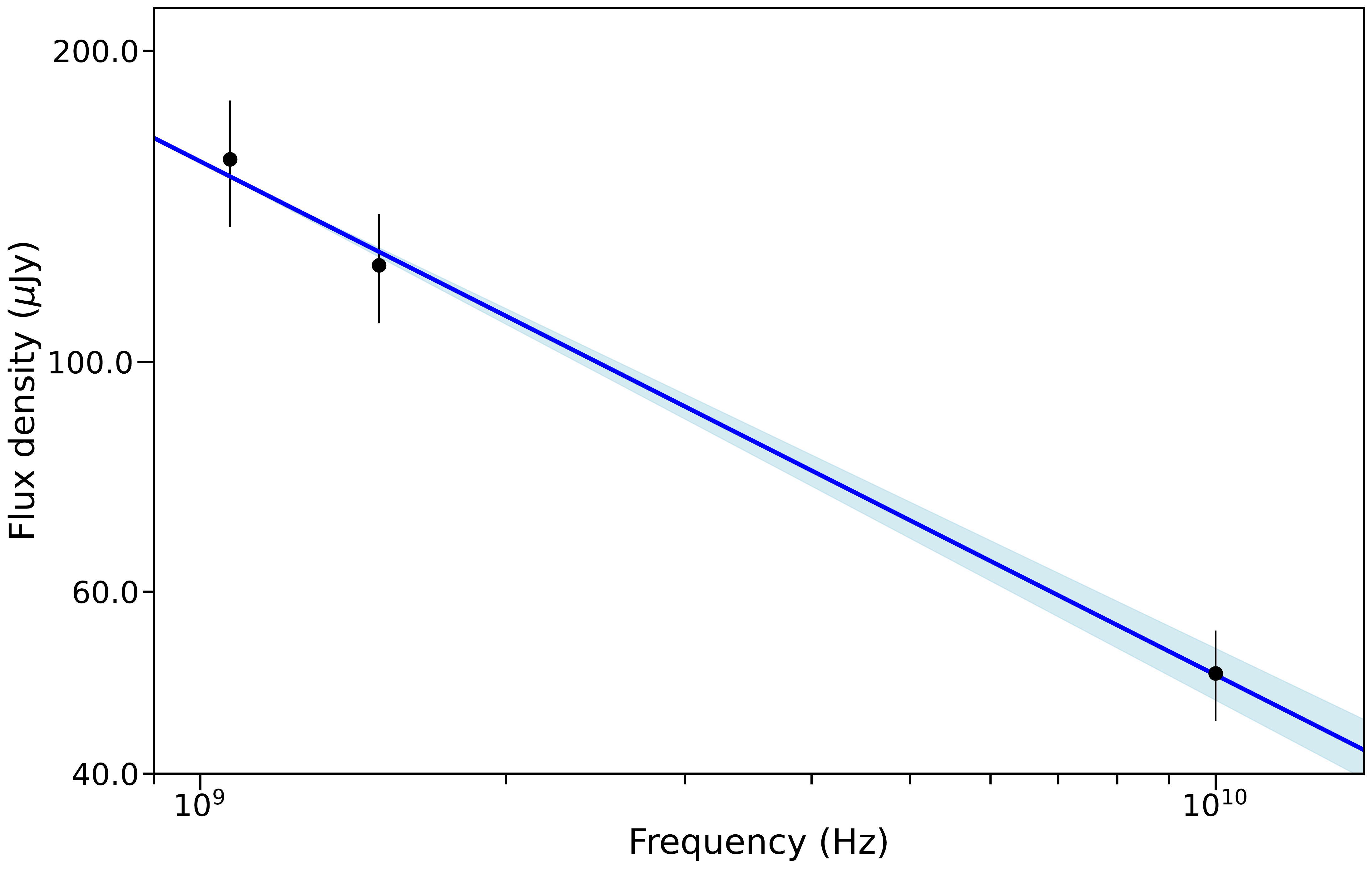}
            \caption[ ]%
            {{\small Radio spectrum from 0.9 to 12\,GHz using observations from MeerKAT and VLA at $\sim$54 days post burst.}}
            \label{fig:mean and std of net44}
        \end{subfigure}
        \caption[ The average and standard deviation of critical parameters ]
        {Various spectra at three separate epochs.} 
        \label{fig:broad_band_spec}
    \end{figure}

There are three epochs where we construct broadband spectra. Figure \ref{fig:mean and std of net24} shows the optical and X-ray data detections (black circle and square), approximately two hours post burst. The blue shaded region denotes the range of possible predicted flux densities under the assumption that $\nu_{\textrm{C}}$ falls between the optical and X-ray bands at the time of the observations. The range is calculated assuming that $\beta_{O-X}$ is between $\beta_{X} + 0.5$ and $\beta_{X}$ i.e. $\nu_{C}$ is at 0.3\,keV and in the r'-band, respectively. It is clear from the blue shaded region that the optical flux density is significantly lower than expected from the synthesised synchrotron spectrum. We calculate that $\beta_{\textrm{O-X}} = -0.13\pm0.02$. Given that $\beta_{X} = -1.0\pm0.2$ at this time, if $\nu_{\textrm{C}}$ falls between the optical and lower end of the X-ray band, we can infer that $\beta_{O-X} \textrm{should be} -0.5\pm0.2$ ($\beta_{O-X} = \beta_{X} + 0.5$), much steeper than our measured $\beta_{\textrm{O-X}}$ \citep{2009ApJ...699.1087V}. We discuss this classification further in section \ref{subsec:dark}. 


Using the detections and upper limits between 20-24\,days post burst (Figure \ref{fig:mean and std of net34}), we can construct a broadband spectrum from 1.3\,GHz to 10\,keV. Around 20\,days, $\beta_{\textrm{10\,GHz-X}} = -0.6$, which differs from $\beta_{X} = -1.0\pm0.1$ at a 4$\sigma$ level indicating that it is likely that $\nu_{\textrm{C}}$ is below the XRT band. We use the value of p (2.0$\pm$0.2) derived from the X-ray spectrum and constrain the location of the cooling break at 20 days to be between 8$\times$10\textsuperscript{15}Hz < $\nu_{\textrm{C}}$ < 8$\times$10\textsuperscript{17}Hz. We also infer the position of the peak of the spectrum and the corresponding flux density: 13$\pm$9\,GHz and 130$\pm$30\,$\mu$Jy, respectively. The resulting broadband spectrum from around 20 days post burst can be well described with a series of three power laws, as shown in Figure \ref{fig:mean and std of net34}: (1) a low frequency steep spectral component, (2) an optically thin branch where $\beta$=-0.5 between $\nu_{\textrm{peak}} < \nu < \nu_{\textrm{C}}$, and (3) the final branch above the cooling break where $\beta$=-1.0$\pm$0.1. The shaded regions denote the uncertainties (for the optically thin and cooling branches) or variations in possible spectral indices (low frequency optically thick branch). 


At 54 days post burst, the broadband radio spectrum is best described by a single power law component (Figure \ref{fig:mean and std of net44}) with a spectral index of $\beta_{\textrm{rad}} = -0.50\pm0.02$, most likely from the optically thin branch of the spectrum below $\nu_{C}$ where p = 2.00$\pm$0.04. This means that by 54 days post burst the peak of the synchrotron spectrum is below 0.9\,GHz.

\subsection{GRB 201216C as a dark GRB}
\label{subsec:dark}

Early time optical observations either placed deep upper limits on any optical emission or obtained very faint detections of the afterglow with respect to the X-ray fluxes, see e.g. Figure \ref{fig:mean and std of net24}. Furthermore, the optical spectrum observed by the VLT was very steep, which provided concrete evidence that GRB 201216C is a dark GRB \citep{2020GCN.29077....1V}. 

From the optical and X-ray light curves, as well as the broadband and optical spectra, we can infer that $\nu_{\textrm{C}}$ is between the optical and X-ray observing bands from 0.05 to $\sim$1\,day post burst. We do not consider the final X-ray data point here because it is a low significance detection. We use this information to place limits on the r'-band flux density if GRB 201216C was not heavily affected by extinction as described in Section \ref{sec:broadband}. By comparing the inferred and measured optical flux densities, we estimate the extinction to be between 5.3 and 8.6\,magnitudes (r'-band); values far in excess of the the galactic extinction contribution: A\textsubscript{R} = 0.12\,mag \citep[E(B-V) = 0.05;][]{1998ApJ...500..525S, 1999PASP..111...63F}.

Using the empirical relations between extinction and neutral hydrogen column density: N\textsubscript{H}$\approx$ 2$\times$10\textsuperscript{21}\,cm\textsuperscript{-2}A\textsubscript{V}, \citep[][]{1995A&A...293..889P, 2009MNRAS.400.2050G}, we can determine whether or not the line of sight hydrogen column density is consistent with the attenuated optical flux densities. For N\textsubscript{H} = 5.07$\times$10\textsuperscript{21}\,cm\textsuperscript{-2}, from X-ray spectra, we estimate $A_{V}$ $\approx$ 3\,mag ($A_{R}$ $\approx$ 2\,mag). Therefore, an additional source of optical extinction is required. From the observed extinction range, we would expect N\textsubscript{H} = 1-3$\times$10\textsuperscript{22}\,cm\textsuperscript{-2}, obtained from the X-ray spectra, which is at least a factor of two higher than the measured N\textsubscript{H} value. 

Given that GRBs occur in regions of high star formation, increased dust in the vicinity of the GRB site is expected, so optically dark GRBs should not be uncommon \citep{2006Natur.441..463F}. Studies of the host galaxies of dark GRBs have shown that the dust distribution is non-uniform, further agreeing with the previous statement that dark GRBs occur in highly obscured regions \citep{2009AJ....138.1690P}. Giant molecular clouds could also be a contributing factor to increased amounts of dust in the vicinity of lGRBs, but would also result in higher measured N\textsubscript{H} values \citep{1987ApJ...319..730S}. 

The fact that we do not observe such a high N\textsubscript{H} may also be a result of the A\textsubscript{V}-N\textsubscript{H} correlation varying from galaxy to galaxy, especially at high redshift (z > 1) where the star formation rate is much higher than in local galaxies. The above calculation assumes a universal A\textsubscript{V}-N\textsubscript{H} relation, and so is not necessarily correct for GRB 201216C's host galaxy. GRB 110709B's optical darkness was similarly under predicted by the measured hydrogen column density \citep{2013A&A...551A.133P, 2013ApJ...767..161Z}. On the other hand, the optical extinction for many GRBs is \textit{over} estimated, again implying a clear deviation from the Galactic and Magellanic Cloud relations \citep{2009AJ....138.1690P, 2011A&A...534A.108K}. 

Of the four other VHE GRBs detected so far, GRBs 190829A and 190114C have also shown increased optical extinction \citep{2021A&A...649A.135C, 2021ApJ...917...95Z}. \citet{2021ApJ...917...95Z} measured an absorption \textit{E(B-V)}=0.757 for GRB 190829A and \citet{2021A&A...649A.135C} obtained \textit{E(B-V)}=0.83 for GRB 190114C. Such high extinctions in the most well studied VHE GRBs could suggest a potential connection between high density/dusty environments and the presence of VHE emission. Dusty environments could result in strong infrared radiation fields following the reprocessing of optical emission. The infrared radiation could be up-scattered to VHE energies in the presence of electrons with sufficiently high Lorentz factors ($\gamma_e \sim 10^6$). Further exploration of this idea is outside the scope of this paper and will be considered in future work.

\section{Discussion}
\label{sec:discussion}

\subsection{Single forward shock model}
\label{subsec:single_FS}

\begin{figure*}
    \centering
    \includegraphics[width = \textwidth]{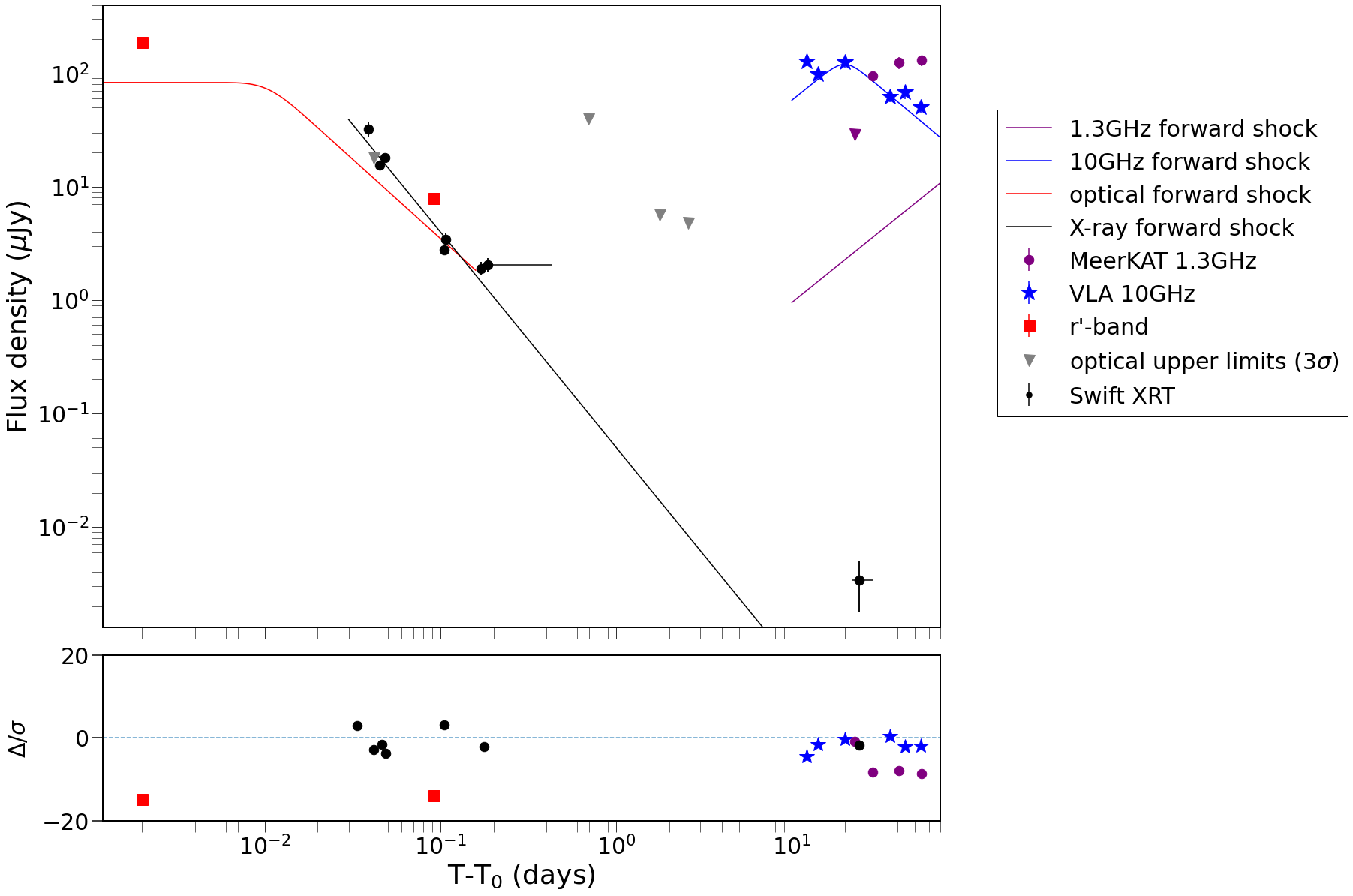}
    \caption{\textit{Top panel:} X-ray, optical, 10\,GHz and 1.3\,GHz data for GRB 201216C. Overlaid is a simple single forward shock model. For the optical light curve, we require an optical extinction between 4.3-4.6\,magnitudes, this is far outside the range of inferred extinction and so we model the light curve with an extinction of 5.3\,magnitudes: the lowest value in our inferred range. \textit{Bottom panel:} the normalised residuals, the ratio of `$\Delta$' (the difference between each observed flux density and the model at that time) to $\sigma$ (the uncertainty on each measured flux density).}
    \label{fig:forward_shock}
\end{figure*}

The most simple description of the multi-frequency data would be a single forward shock which is produced as the jet decelerates in the circumburst medium. Using our constraints on the positions of the break frequencies and the peak flux from our broadband spectral considerations (Section \ref{fig:broad_band_spec}), combined with well established analytical afterglow models \citep[e.g.][]{2002ApJ...568..820G}, we can determine if a single forward shock describes our data well. 

In Section \ref{sec:broadband}, we constrained the location of $\nu_{\textrm{C}}$ at 20 days post burst to be between 8$\times$10\textsuperscript{15} and 8$\times$10\textsuperscript{17}\,Hz. Given that we observe no statistically significant breaks in the light curve before day 20, we can assume that until at least day 20, the X-ray emission must originate from above $\nu_{\textrm{C}}$. For an X-ray decay of $\alpha_{X} = -1.9\pm0.2$, using the binning shown in Figure \ref{fig:lc}, we obtain p = 3.2$\pm$0.3, which is steeper than p derived from the X-ray spectra (2.0$\pm$0.4) at nearly 2$\sigma$. These p values are independent of the circumburst medium. Despite the light curve slope being same, the movement of $\nu_{\textrm{C}}$ changes depending on the environment. In a stellar wind environment $\nu_{\textrm{C}}\propto t^{\frac{1}{2}}$ and in a homogeneous medium $\nu_{\textrm{C}} \propto t^{-\frac{1}{2}}$. Given the inference that $\nu_{\textrm{C}}$ is only just below the XRT band at day 20; if the jet was propagating through a homogeneous environment, we would expect to observe a break in the X-ray light curve due to $\nu_{\textrm{C}}$ at some time before day 20. Therefore, we conclude that the XRT emission is most likely a result of a stellar-wind environment where $\nu_{\textrm{C}}$ moves from lower to higher frequencies with time, i.e. towards the XRT band.

\begin{figure}
    \centering
    \includegraphics[width = \columnwidth]{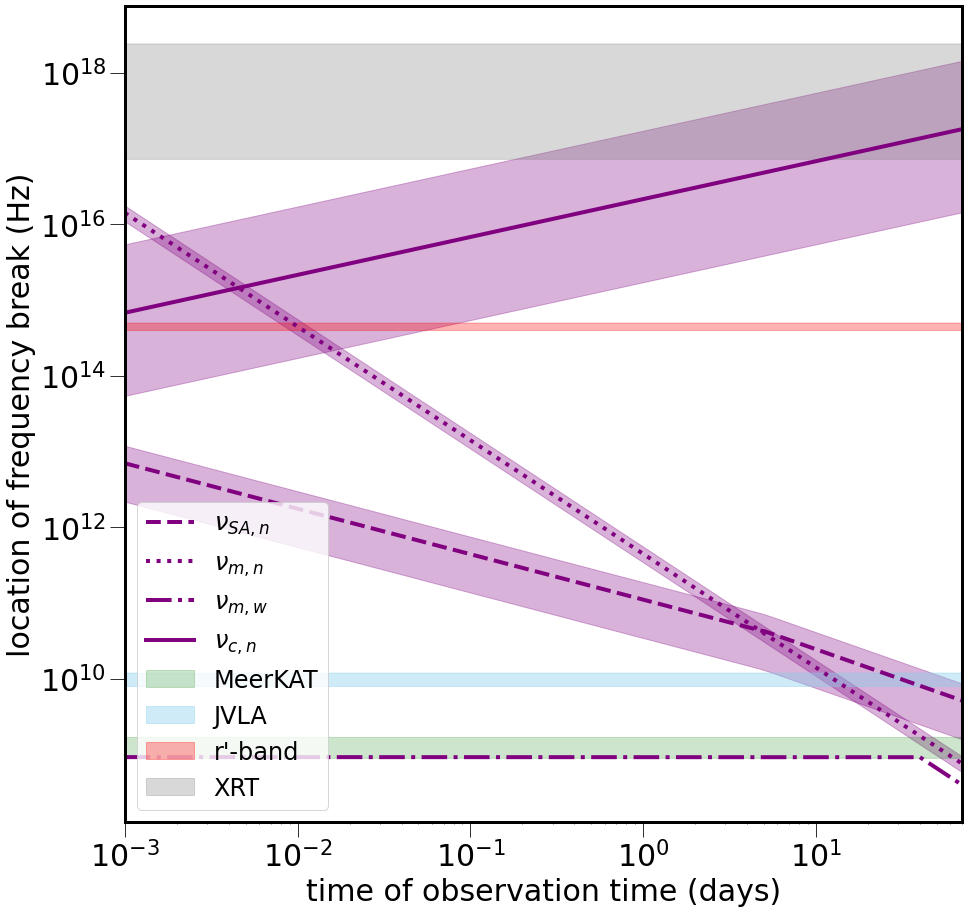}
    \caption{A plot showing the evolution of the break frequencies for both the narrow and wide components from \citet{2005ApJ...626..966P}'s two component afterglow model, adapted for a stellar-wind environment applied to our dataset. The purple dashed, dotted and solid region denotes the movement of $\nu_{\textrm{SA}}$, $\nu_{\textrm{m}}$ and $\nu_{\textrm{C}}$ from the narrow jet, respectively. The dotted-dashed line shows the evolution of $\nu_{\textrm{m}}$ from the wide component viewed off-axis. The purple shaded regions denote the uncertainties in the location of each frequency break. Overlaid in green, blue, red and grey are the MeerKAT, VLA, optical and XRT observing bands, respectively. }
    \label{fig:peak_two}
\end{figure}

\begin{figure}
    \centering
    \includegraphics[width = \columnwidth
    ]{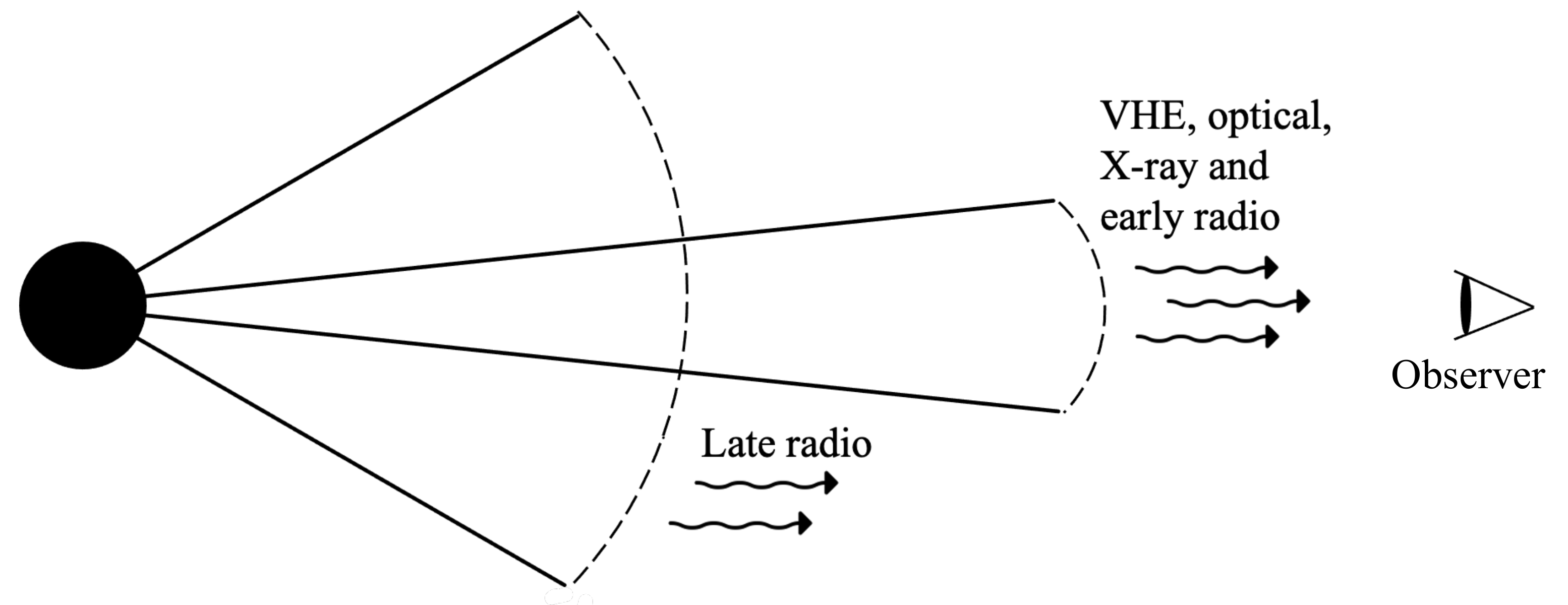}
    \caption{A schematic of the geometry of VHE GRB 201216C. The narrow core component is the origin of the VHE, optical, X-ray and early ($<$ 25 days) radio data. The wider cocoon component produces the late-time radio emission. }
    \label{fig:diagram}
\end{figure}

The model X-ray light curve is shown as the black line in Figure \ref{fig:forward_shock}. From either the X-ray light curve or spectra, it is impossible to determine the location of $\nu_{\textrm{C}}$ with respect to the 0.3-10\,keV band. However, when combining the two with the broadband spectrum at 20 days, we have good evidence that the jet is propagating through a stellar wind environment. Figure \ref{fig:peak_two} shows the movement of $\nu_{\textrm{C}}$ with time according to our stellar wind model (solid purple line). The shaded purple region around the line represents the uncertainty on the location of $\nu_{\textrm{C}}$ at a given time, derived from the broadband spectrum constructed around day 20 (Figure \ref{fig:mean and std of net34}). 

The intrinsic brightness of the optical emission is heavily absorbed, but we can assume that the decay rate ($t^{-0.83\pm0.01}$) observed is unaffected by the material causing the absorption. For a wind environment, such a decay is too shallow to be produced by optically thin synchrotron in which the r'-band is below $\nu_{\textrm{C}}$. The decay, $\alpha_{\textrm{opt}}$, should follow $-2.0 < \alpha_{\textrm{opt}} < -1.3$, using an optically thin decay in a stellar wind environment for 2 $<$ p $<$ 3. The shallower decay that we observe could be a result of $\nu_{\textrm{m}}$ passing through the optical observing band within a few hours of the burst (as previously mentioned in Section \ref{subsubsec:opt_lc}). We use this to constrain the position of $\nu_{m}$ at early times as the purple dotted line in Figure \ref{fig:peak_two}. The red line in Figure \ref{fig:forward_shock} shows the model light curve for a forward shock where $\nu_{\textrm{m}}$ causes the break around 0.01\,days. In order to best fit to the optical data, we consider $\sim$5.3\,magnitudes of optical extinction, the lowest value in the extinction range calculated in Section \ref{subsec:dark}. Figure \ref{fig:forward_shock} shows that our forward shock model does not fit the optical data well.

The similar flux densities, to within an order of magnitude, are observed by both XRT before 0.1\,days post burst and at the beginning of the VLA observing campaign. We estimate that the peak of the spectrum should be in the optical band around 0.01\,days. Figure \ref{fig:forward_shock} shows that at 0.1\,days, the most optical and X-ray flux densities are approximately the same. However, when we consider the extinction calculated in Section \ref{subsec:dark}, it is clear that F\textsubscript{$\nu$, max} should be significantly higher at this time. The VLA observations, which start at day 20 and show flux densities similar to that measured by XRT 0.04 days post burst, show evidence of the spectral peak passing through the 10\,GHz band, meaning that F\textsubscript{$\nu$, max} must have decayed from 0.01 to 20 days post burst to show similar X-ray and radio flux densities. If we assumed an ISM environment, F\textsubscript{$\nu$, max} would be constant with time and would not describe the data at all. Instead, this can be attributed to either a jet break or a stellar-wind environment. A jet break can cause F\textsubscript{$\nu$, max} to decay rapidly as a result of the jet expanding laterally \citep[F\textsubscript{$\nu$, max} $\propto t^{-1}$, ][]{Sari1999jets}. A stellar wind environment means that F\textsubscript{$\nu$, max} will decay with time as F\textsubscript{$\nu$, max}$\propto t^{-0.5}$ when $\nu_{\textrm{m}}$ > $\nu_{\textrm{SA}}$, and F\textsubscript{$\nu$, max}$\propto t^{-\frac{4p+1}{2(p+4)}}$ ($t^{-(0.75-0.93)}$ for 2$<$p$<$3) when $\nu_{\textrm{SA}}$ > $\nu_{\textrm{m}}$ \citep{2002ApJ...568..820G}.

The VLA in-band spectral indices of the first three observations are steep (see Table \ref{tab:rad_obs}). Such values further imply that the peak of the synchrotron spectrum is above 12\,GHz until at least 20 days post burst (see the upper panel of Figure \ref{fig:broad_band_spec}). As explained in Section \ref{subsubsec:radio}, if the peak of the spectrum was caused by $\nu_{\textrm{m}}$ then we would expect to measure a spectral index of $\beta = 1/3$. The steepness of the spectral indices implies that either $\nu_{\textrm{SA}}$ and $\nu_{\textrm{m}}$ or just $\nu_{\textrm{SA}}$ must be above the VLA observing band. The third VLA epoch is more consistent with a flat spectrum, i.e. $\beta = 1/3$ or the peak of the spectrum being at the observing frequency. Combined with the decaying flux density of the final three epochs, we can conclude that the peak of the synchrotron spectrum has passed through the 8-12\,GHz band between 20 and 36 days post burst. The most simple scenario for the decay is if only one spectral break is above the VLA band, which moves towards lower frequencies with time, causing the emission above the break to decay with time. In the regime where $\nu_{\textrm{m}}$ < $\nu_{\textrm{SA}}$, $\nu_{\textrm{SA}} \propto t^{-\frac{3(p+2)}{2(p+4)}}$ \citep[$t^{-(1.0-1.1)}$ for 2$<$p$<$3,][]{2002ApJ...568..820G}. Therefore, we conclude that the peak of the spectrum is caused by $\nu_{SA}$.

At 10\,GHz, the forward shock model would show a broken power law, with a rise following $t^{\frac{5}{4}}$ to a peak around day 20 followed by a decay of $\alpha_{\textrm{10\,GHz}} = \frac{(1-3p)}{4}$ ($t^{-(1.3-2.0)}$ for 2 $<$ p $<$ 3) as $\nu_{\textrm{SA}}$ moves through the observing band \citep{2002ApJ...568..820G}. The theoretical 10\,GHz light curve is shown (where p = 2.0) with the blue line in Figure \ref{fig:forward_shock}. The optically thick to thin transition provides a reasonable fit to the 10\,GHz light curve. We note that the variability due to weak scintillation increases the residuals as shown in the lower panel of Figure \ref{fig:forward_shock}.

Figure \ref{fig:mean and std of net44} shows that by 54 days post burst, the 1.3 and 10\,GHz light curves are on the same, optically thin branch, of the synchrotron spectrum, with p = 2.00$\pm$0.05. The MeerKAT in-band spectral index is also similar at day 41 post burst indicating that by 41 days post burst, 1.3\,GHz is above $\nu_{\textrm{SA}}$ and $\nu_{\textrm{m}}$. In order for $\nu_{\textrm{SA}}$ to be below the MeerKAT band at 41 days post burst, we required $\nu_{\textrm{SA}} \propto t^{-3.5}$, which is significantly faster than the theoretical movement where for p $\approx$ 2, $\nu_{\textrm{SA}} \propto t^{-\frac{3(p+2)}{2(p+4)}} = t^{-0.8}$ \citep{2002ApJ...568..820G}. The measured $\nu_{\textrm{SA}}$ movement is unphysical when compared to analytical models.

The unphysical movement for $\nu_{\textrm{SA}}$ complicates the 1.3\,GHz model light curve. Using  $\nu_{\textrm{SA}} \propto t^{-0.8}$, the model light curve would consist of a single power law component with $\alpha_{\textrm{1.3\,GHz}} = \frac{5}{4}$. The light curve would turn over at 200 days as a result of $\nu_{\textrm{SA}}$ entering the observing band if $\nu_{\textrm{SA}}$ moved as given in \citet{2002ApJ...568..820G}. During the rise, we would expect a spectral index $\beta_{\textsubscript{1.3\,GHz}} = \frac{5}{2}$, not $\sim$-1 as measured.  Furthermore, the peak flux density, according to our single shock model would continue to decay as F\textsubscript{$\nu$, max}$\propto t^{-\frac{4p+1}{2(p+4)}} = t^{-0.75}$ for p $=$ 2, so by the time $\nu_{\textrm{SA}}$ (the peak of the spectrum) reaches the MeerKAT band, F\textsubscript{$\nu$, max}$\approx$20$\mu$Jy, which is a factor of 7 fainter than the observed 1.3\,GHz flux density. When compared to the MeerKAT data points (purple circles and downwards facing triangle), the 1.3\,GHz model light curve (the purple line) in Figure \ref{fig:forward_shock} show a clear deviation away from the forward shock model. At all times, the model light curve falls far below the observed data. Even if we assume the unphysical movement of $\nu_{\textrm{SA}}$, F\textsubscript{$\nu$, max} still decays with time meaning that the modelled 1.3\,GHz light curve is predicted to be much fainter than the observed emission.

The late-time change in the evolution of F\textsubscript{$\nu$, max} and $\nu_{\textrm{SA}}$ could be a result of a change in the circumburst environment. A varying circumburst density distribution, i.e. a deviation from $\rho \propto r^{-2}$, where $\rho$ and r are the density and radius for the burst site, could occur as a result of the progenitor star having fluctuating mass loss rates towards the end of its life. In order to reproduce the observations, we require the circumburst environment to change from $\rho \propto r^{-2}$ to $\rho \propto r^{0}$: an homogeneous environment. Such a change seems unlikely to reflect the mass loss history of the progenitor. Furthermore, this cannot explain the MeerKAT light curve or the discrepancy in p derived from the X-ray light curve and spectra.

In conclusion, Figure \ref{fig:forward_shock} shows the results of a single forward shock component model overlaid on the X-ray, optical, 10\,GHz and 1.3\,GHz light curves. We do not use the \textit{e}-MERLIN 5\,GHz data points in our model as they are heavily affected by DISS. The bottom panel of Figure \ref{fig:forward_shock} shows the normalised residual values for our forward shock model with respect to the data. We are able to reproduce the X-ray and VLA light curves reasonably well. The inter-observation variability increases the residuals of the VLA data with respect to the model. However, it is clear from the large residuals for optical and MeerKAT light curves that a single forward shock model is not a good fit. In a single shock scenario, we expect any variation in F\textsubscript{$\nu$, max} to be dictated by the movement of $\nu_{\textrm{SA}}$ and $\nu_{\textrm{m}}$. We acknowledge that our understanding of how F\textsubscript{$\nu$, max} evolves with time early on is poorly constrained because of the faint optical emission. By inferring the range of optical flux densities from the X-ray data, we know that we require a steeper F\textsubscript{$\nu$, max} decay than the afterglow models provide \citep{2002ApJ...568..820G}. On the other hand,  by the time the radio campaign begins, F\textsubscript{$\nu$, max} appears to be constant in time. The rapid decay of F\textsubscript{$\nu$, max} until day $\sim$20 followed by a transition to a constant F\textsubscript{$\nu$, max} for the rest of the observing campaign is too complex to be attributed to a single jet component.

\subsection{Multiple shock component model}

It is possible that the discrepancies the light curves could be explained with the addition of an extra shock component. Unfortunately, the light curves in any one observing band are too sparse to search for reverse shock emission. For example, optically thin reverse shock emission decays far steeper than the optical light curve. However, we note that two data points are not constraining enough to fully eliminate the possibility of reverse shock contribution. Unfortunately, because our first data point from \textit{e}-MERLIN at 5\,days post burst is dominated by RISS, we do not know the intrinsic flux density value at this time and so cannot determine if there is any reverse shock contribution. The full 5\,GHz flux density in the absence of RISS (double/half the observed flux density assuming order of unity variability) is not constraining enough to confirm or reject the presence of reverse shock emission.

The MeerKAT and VLA bands are less affected by ISS and therefore are more appropriate to search for reverse shock emission. We apply a similar methodology to that in \citet{2020MNRAS.496.3326R} in order to determine whether the reverse shock makes a significant contribution to our 1.3\,GHz light curve despite not being detectable at 10\,GHz. If the reverse shock has faded at 10\,GHz by 12\,days post burst, then we can determine that the 29\,days, whilst the peak of the reverse shock might be in the MeerKAT observing band, F\textsubscript{$\nu$, \textrm{max}} corresponding to the reverse shock will be significantly fainter than the observed MeerKAT flux densities at this time. The strongest evidence against reverse shock emission in the MeerKAT band is that the sharp rise between 22 and 29\,days post burst, inconsistent with the reverse shock scenario. 

An additional forward shock component could instead explain the large residuals between the data and the single shock model for the later observations at 1.3\,GHz. In order to determine whether an additional forward shock could explain the discrepancies found in Section \ref{subsec:single_FS}, we use the two component jet model presented in \citet{2005ApJ...626..966P} adapted for a stellar wind environment \citep{1999ApJ...520L..29C} to better interpret our data, similar to what was applied to GRB\,130427A by \citet{2014MNRAS.444.3151V}. Structured jets encompass a broad range of geometries and multiple jet components have been invoked in previous GRB afterglow data sets \citep[e.g.][]{2005A&A...440..477R, 2008Natur.455..183R}. The afterglow from gravitational wave event GW 170817 was inferred to have an ultra-relativistic core surrounded by lower velocity wings \citep[][]{2018ApJ...856L..18M}. In GRB 080319B, the presence of two distinct outflow components was inferred, but the wider component was still more collimated than what we infer for our narrow jet \citep{2008Natur.455..183R}. The presence of a wider component is strongly supported by simulations \citep[e.g.][]{2007ApJ...665..569M}, although observations of such a component span a broad range of energetics and opening angles.

For our data on GRB 201216C, we consider a narrow, ultra-relativistic jet launched at a Lorentz factor, $\Gamma_{0, n} > 100$, like with the single shock scenario, but with the addition of a wider outflow with $\Gamma_{w, 0} < 10$ (the subscripts \textit{n} and \textit{w} refer to \textit{narrow} and \textit{wide}, respectively). It is possible for the wider outflow to be non-relativistic, but we find that a relativistic outflow is more likely given the high luminosity and light curve behaviour. We can rule out the possibility of supernovae emission as the origin of the late time radio detections: the radio luminosities are an order of magnitude higher than the next most luminous type Ib/c broad-line supernova \citep{2021ApJ...908...75B}.

Each component is uniform within some opening angle $\theta_{j, n}$, $\theta_{j, w}$. The two components do not interact with each other. In the event of a jet break, we only consider edge effects, not lateral spreading. An on-axis observer would see that the narrow jet dominates at early times. The wider component becomes visible only in systems with favourable geometries and energetics. Figure \ref{fig:diagram} shows the geometry of such a system, which undergoes a jet break early on ($\sim$0.05\,days) allowing the wider outflow to be visible later on.

\begin{figure*}
    \centering
    \includegraphics[width=\textwidth]{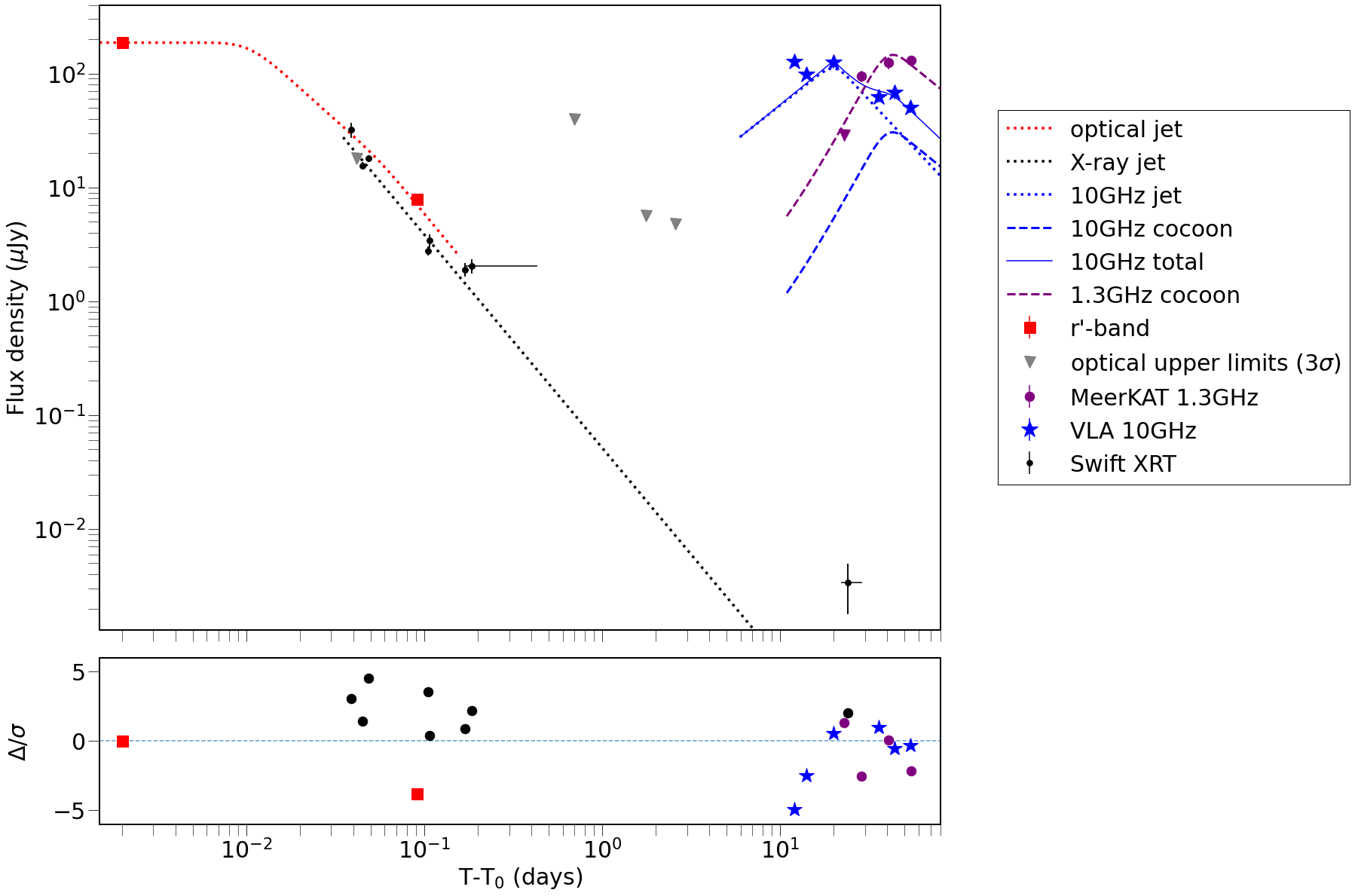}
    \caption{\textit{Top panel}: X-ray, optical and radio data. Overlaid are the model light curves corresponding to the two jet component scenario. In this scenario, we require a range of optical extinction between 7.4-8.0\,magnitudes. \textit{Bottom panel}: the normalised residuals, the ratio of `$\Delta$' (the difference between each observed flux density and the model at that time) to $\sigma$ (the uncertainty on each measured flux density). }
    \label{fig:two_shocks}
\end{figure*}

\subsubsection{Narrow component}
\label{subsub:narrow}

In Section \ref{subsec:single_FS}, we noted that the evolution of F\textsubscript{$\nu$, max} is inconsistent with a single forward shock. We can explain a steep decay of F\textsubscript{$\nu$,max} as well as the X-ray light curve if the narrow component undergoes a jet break before the first X-ray data points at around 0.05\,days post burst. As with the single jet scenario, the X-ray light curve follows a single power law decay. For a jet break scenario in a wind environment, we obtain p = 2.2$\pm$0.2, which agrees with p from the X-ray spectra (p = 2.0$\pm$0.4) at a 1$\sigma$ level, slightly better than in the single shock scenario where p = 3.2$\pm$0.4 (Section \ref{subsubsec:X-raylc}).

At 2\,hours post burst, we infer from from Figures \ref{fig:mean and std of net24} and \ref{fig:peak_two} that 3 $\lesssim$ F\textsubscript{$\nu$, max} $\lesssim$ 15\,mJy. By 20\,days, F\textsubscript{$\nu$, max}$=$ 0.13$\pm$0.03\,mJy (Figure \ref{fig:mean and std of net34}). We find that the best explanation of the steep decay of F\textsubscript{$\nu$,max} is a result of a combination of the jet break and the wind environment. The combination creates a steeper decay of F\textsubscript{$\nu$, max} compared to a stellar wind-only decay as used in the single shock scenario. The steepening of the light curves due to a jet break is a correction factor of 0.75 \citep{1998ApJ...503..314P}. Once we consider the optical extinction, we can match the inferred optical flux density from our model with observed optical flux densities. We use the spectral evolution to explain the optical detections here as in Section \ref{subsec:single_FS}: $\nu_{\textrm{m}}$ passing through the band resulting in an apparent flattening of the optical light curve. The dotted red line in Figure \ref{fig:two_shocks} shows a broken power law where the break is a result of $\nu_{\textrm{m}}$ passing through the band, in addition to the jet break at 0.05\,days. The optical emission is best fit with an extinction range between of 7.4-8.0\,magnitudes which is within the extinction range inferred from the X-ray observations (5.3-8.6\,magnitudes). For the multiple component jet model, we require a larger extinction value to fit the optical light curve well because at earlier times F\textsubscript{$\nu$,max} is brightest compared to the single shock model. The lower panel of Figure \ref{fig:two_shocks} shows that this optical model fits the data significantly better, compared to in Figure \ref{fig:forward_shock} where in order to get the best fit the assumed extinction was outside the inferred range.

In terms of the radio data, the narrow jet dominates the early VLA light curve. The narrow jet contribution to the measured 10\,GHz emission is shown with the blue dashed line in Figure \ref{fig:two_shocks}. The 10\,GHz narrow component light curve follows a sharp rise, $\alpha_{\textrm{10\,GHz}} = \frac{5}{4}$, followed by a decay as $\alpha_{\textrm{10\,GHz}} =-\frac{3p}{4} = -1.7$ for p = 2.2.  After the peak, the 10\,GHz emission decays rapidly giving way to a wider component. Both the rise and the decay are steeper than in the single shock scenario because by the first VLA epoch the narrow jet component has already undergone a jet break. The first two VLA spectra show that the narrow component is self-absorbed. Therefore, the 1.3\,GHz narrow jet would peak at around 2$\mu$Jy, and therefore we do not show the 1.3\,GHz contribution to the narrow component. 
 
Despite GRB 201216C having a VHE counterpart, our modelling of the afterglow considers only synchrotron emission. Unfortunately, it is not possible to model the SSC emission for GRB 201216C given the sparse sampling in the X-ray energy range and the lack of public VHE data. However, the low value of $\epsilon_{B}$ we derived from our afterglow modelling of the narrow jet implies that the narrow jet is in a regime where SSC cooling dominates over synchrotron cooling at least at early times when there is a high fraction of electron energy lost due to radiation \citep{2001ApJ...548..787S}. SSC modelling is outside the scope of this work, however, in future studies, we plan to use new tools such as detailed modelling code by \citet{2021MNRAS.504..528J} which consider SSC cooling. 

Our optical light curve, VLA in-band spectral indices and broadband spectrum at 20 days, allow us to constrain the locations of $\nu_{\textrm{SA}}$, $\nu_{\textrm{m}}$ and $\nu_{\textrm{C}}$ for the narrow jet. We combine analytical models for the movement of the frequency breaks (Figure \ref{fig:peak_two}) along with the decay of F\textsubscript{$\nu$, max}, and extract physical parameters from this dataset: E\textsubscript{K,\footnotesize{ISO},n}, A\textsubscript{\footnotesize$\ast $}, $\epsilon_{\textrm{e}}$ and $\epsilon_{\textrm{B}}$ (assuming p=2.2 from the X-ray data). These ranges of derived values for those paramters are given in Table \ref{tab:params}. We also give the opening angle of the narrow component which corresponds to a jet break at 0.05 days. From the jet break, we obtain an opening angle of $\sim$1-9$^{\circ}$ \citep{2000ApJ...536..195C} which corresponds to a beaming corrected opening energy between (0.01-20)$\times$10\textsuperscript{50}\,erg.


The stellar wind environment we infer from this data set is characterised by the parameter A\textsubscript{*}. \citet{1999ApJ...520L..29C} relates A\textsubscript{*} to the density profile $\rho = Ar^{-2}$ where A = $\dot{M}$/$4\pi v_{w}$ = 5$\times$10\textsuperscript{11}A\textsubscript{*} g\,cm\textsuperscript{-1}. If $v_{w}$ = 1000\,km\,s\textsuperscript{-1}, we obtain a progenitor mass loss rate ($\dot{\textrm{M}}$) of (0.6-200)$\times$10\textsuperscript{-5}M\textsubscript{\(\odot\)}/yr. Winds of massive stars are heavily dependent on metallicity and GRBs are expected to occur in low metallicity environments. If the metallicity is too high, the mass loss rate would also be too high resulting an increased loss of angular momentum which inhibits the formation of the GRB. Furthermore, the stellar wind is expected to have a non-spherical distribution with the majority of the material concentrated around the equator. Therefore, at the poles, we expect a less distinct stellar wind profile compared to an equatorial view. The inferred range of mass loss rate for GRB 201216C from our two component model is within the range of expected values \citep{2005A&A...442..587V, 2018ApJ...858..115A}.

We can use the $\dot{\textrm{M}}$ range derived from the afterglow modelling to infer limits on the progenitor mass \citep{1989A&A...210...93L, 2016ApJ...833..133T, 2017MNRAS.470.3970Y}. The upper end of our $\dot{\textrm{M}}$ range is pushing the boundaries for the progenitor to be a Wolf-Rayet star, independent of metallicity. Low metallicities alongside with high stellar masses would be required to begin to reach such high mass losses, combined with inciting gravity waves \citep{2018MNRAS.476.1853F} or if the star is reaching the Eddington luminosity \citep{1994A&A...290..819L}. At the lower end of our $\dot{\textrm{M}}$ range, the progenitor could be a Wolf-Rayet star of 10\.M\textsubscript{\(\odot\)} at solar metallicity or even 20-25\,M\textsubscript{\(\odot\)} at sub-solar metallicity. Since GRBs tend to occur in low metallicity environments, the progenitor mass is likely to be between 12-25\,M\textsubscript{\(\odot\)}. 

\begin{table}
    \centering
    \begin{tabular}{c|c|c}
        \hline
        Parameter & Narrow & Wide \\
        \hline
        E\textsubscript{ISO,K} (erg) & (0.6-10)$\times$10\textsuperscript{52} & (0.02-50)$\times$10\textsuperscript{48} \\
        A\textsubscript{*} & 0.6-200 & 0.6-200\\
        $\epsilon_e$ & 0.04-0.1 &  0.1\\
        $\epsilon_B$ & 5$\times$10\textsuperscript{-8} - 4$\times$10\textsuperscript{-3} & 0.01\\
        p & 2.0-2.4 & 2.0 \\
        $\theta_{j}$ & 1-9$^{\circ}$ & - \\
        \hline
    \end{tabular}
    \caption{The physical parameters extracted from our data set using \citet{2005ApJ...626..966P}'s two component model. For the wider outflow, we have reduced coverage and therefore assume the same range of values for A\textsubscript{$\ast $} and use fiducial values of $\epsilon_{e} = 0.1$ and  $\epsilon_{B} = 0.01$. We constrain \textit{p} for the wider outflow from the MeerKAT-VLA spectral index measured at 54\,days post burst.} 
    \label{tab:params}
\end{table}

\subsubsection{Wider outflow}
\label{subsubsec:wide}

The decay of the narrow component allows us to detect the wider outflow. The wider outflow is the origin of the observed 1.3\,GHz emission: the MeerKAT light curve shows a sharp rise from 22 days, best described as emission from such a second jet component. Using the behaviour of $\nu_{\textrm{m}}$ and F\textsubscript{$\nu$, max} for off-axis jets in a stellar wind environment \citep{2005ApJ...626..966P, 1999ApJ...520L..29C}, we can constrain the time at which the outflow comes into our line of sight (t\textsubscript{on}) from both the light curve and the spectra. For t< t\textsubscript{on}, $\nu_{\textrm{m}} \propto t^{0}$, and for t > t\textsubscript{on},  $\nu_{\textrm{m}} \propto t^{-\frac{3}{2}}$. The dotted-dashed purple line at the bottom of Figure \ref{fig:peak_two} shows the movement of $\nu_{\textrm{m}}$ in the wide jet component. Comparison of the spectral index measurements between day 28 ($\beta_{1.3\,GHz} > -0.3$) and 41 ($\beta_{1.3\,GHz} = -1.1\pm0.6$, see also Figure \ref{fig:mean and std of net44}) shows that there is some movement of $\nu_{\textrm{m}}$ between the two epochs. Therefore t\textsubscript{on} must occur before day 41. We can further constrain the deceleration time by looking at the evolution of F\textsubscript{$\nu$, max}: for t $<$ t\textsubscript{on}, F\textsubscript{$\nu$, max} $\propto t^{3}$, for t $>$ t\textsubscript{on}, F\textsubscript{$\nu$, max} $\propto t^{-\frac{1}{2}}$, we place t\textsubscript{on} at $\sim$40\,days \citep{2005ApJ...626..966P}.

Unlike for the narrow component, we do not have broadband observations, and as a result we are unable to perform the same detailed modelling as presented in section \ref{subsub:narrow}. We are able to determine the range of kinetic energies by assuming the same range of A\textsubscript{$\ast $} as for the narrow component and assuming that $\epsilon_{e}$ and $\epsilon_{B}$ are 0.1 and 0.01, respectively. We can calculate E\textsubscript{K, ISO, w} in the wider outflow to be (0.02-50)$\times$10\textsuperscript{48}erg. These values are also summarised in Table \ref{tab:params}. We find that the isotropic equivalent kinetic energy present in the wide outflow is two and seven orders of magnitude lower than in the narrow component for the same stellar wind profile. We note that the inferred range of kinetic energies for the cocoon is dependent on the assumed values of $\epsilon_{e}$ and $\epsilon_{B}$. With the assumed microphysical parameters, the kinetic energy of the cocoon can be considered as mildly to non-relativistic when compared to other radio transients, similar to radio-detected supernovae \citep[e.g. figure 5 of ][]{2020ApJ...895L..23C}. However the rapid rise and high luminosity are inconsistent with type Ib/c supernovae (those associated with long GRBs) and so the outflow is more likely to be mildly relativistic.  

Figure \ref{fig:two_shocks} shows the wide jet contribution to the 1.3\,GHz and 10\,GHz light curves as the purple and blue dashed lines, respectively. We also show the total 10\,GHz model from both jet components as the solid blue line. We expect the wide component of the jet to make some contribution of the total X-ray flux observed by XRT at the time of our final observation which would make the model closer to the observed flux density. We are unable to quantify the contribution of the wide jet component to the total X-ray flux as we do not know the location of $\nu_{\textrm{C}}$ with respect to the XRT observing band. 


In comparison to our single shock scenario, the normalised residual values are much lower denoting a better fit (shown in the lower panel of Figure \ref{fig:two_shocks}). There are still some increased residual values early on in the 10\,GHz light curve, although this may be due to weak interstellar scintillation which we cannot model. A much wider outflow makes for a much better fit to the MeerKAT light curve as well as the later VLA data points. We can better quantify whether the more complex, multiple jet component model is a better fit compared to the single forward shock model by performing an F-test. The single jet model has four parameters and the two component model has only one additional free parameter originating from the wider component, the kinetic energy. The results show that our more complex model is favoured at greater than 4$\sigma$ significance.



\section{Conclusions}

We have presented multi-wavelength observations of the fifth VHE GRB. Spectra of the host galaxy confirmed GRB 201216C to be the highest redshift VHE GRB so far, close to the theoretical distance limit beyond which pair-production due to the EBL would prevent the detection of any VHE photons. The faint early optical detections defined this event as a dark GRB where we infer that the optical emission is attenuated by at least 5\,magnitudes. Such attenuation is at least two magnitudes greater than that derived from galactic A\textsubscript{V}-N\textsubscript{H} relations. Such high extinction could be due to a high dust density in the vicinity of the lGRB, for instance if the stellar progenitor did not travel far from its formation site in the centre of a giant molecular cloud. At radio frequencies, we obtained MeerKAT (1.3\,GHz), \textit{e}-MERLIN (5\,GHz) and VLA (10\,GHz) observations covering 5 to 55 days after the burst. Our \textit{e}-MERLIN epochs show evidence of DISS, allowing us to place emitting region size constraints at 29\,days post burst of $<$1$\times$10\textsuperscript{17}\,cm \citep{1997NewA....2..449G}. 

We interpret the data set as a whole using two possible scenarios. The first is a single forward shock, but this scenario does not explain the data well because F\textsubscript{$\nu$, max} decays in a non constant manner. In the first 20 days after the burst, we required a rapid decay however, later on, we needed F\textsubscript{$\nu$, max} to be constant. A varying F\textsubscript{$\nu$, max} decay could be a result of the jet propagating through a highly variable circumburst medium. However, one would also expect the movement of the frequency breaks to vary which is inconsistent with the inferred break evolution. We require $\nu_{\textrm{SA}}$ to move to lower frequencies at an unphysical rate to explain the late-time broadband spectra, and even if such a movement were possible, the resulting MeerKAT light curve would still rise too quickly. Both the variation in F\textsubscript{$\nu$, max} and the evolution of $\nu_{\textrm{SA}}$ are inconsistent with a forward shock model.

Instead, we suggest a second scenario: a jet-cocoon geometry where the earlier emission is dominated by a narrow ultra-relativistic jet which undergoes a jet break at 0.1\,days. The later time emission is dominated by a wide-angled, slower moving outflow: a cocoon. We find that the additional component allows for fixing the problems regarding the rapid decay of F\textsubscript{$\nu$, max} early on, produced by the jet break, and then the flattening occurs as the radio observations are now viewing the cocoon as a separate synchrotron component. The cocoon also addresses the unphysical movement of $\nu_{\textrm{SA}}$. We find that the jet-cocoon scenario shows the afterglow to be moving through a stellar wind environment of a density similar to that modelled for massive stellar-winds with energies of (0.6-10)$\times$10\textsuperscript{52} and (0.02-50)$\times$10\textsuperscript{48}\,erg for the jet and cocoon respectively. We constrain on the opening angles of the ultra-relativistic jet to be $1-9^{\circ}$. Deeper, more late-time observations are required moving forward in order to better understand and constrain cocoon emission in lGRB events.

\section*{Acknowledgements}
The authors thank the referee for their helpful comments. L. R. acknowledges the support given by the Science and Technology Facilities Council through an STFC studentship. D. R. A.-D. is supported by the Stavros Niarchos Foundation (SNF) and the Hellenic Foundation for Research and Innovation (H.F.R.I.) under the 2\textsuperscript{nd} Call of ``Science and Society'' Action Always strive for excellence - ``Theodoros Papazoglou'' (Project Number: 01431). This work made use of data supplied by the UK Swift Science Data Centre at the University of Leicester. The MeerKAT telescope is operated by the South African Radio Astronomy Observatory, which is a facility of the National Research Foundation, an agency of the Department of Science and Innovation. \textit{e}-MERLIN is a National Facility operated by the University of Manchester at Jodrell Bank Observatory on behalf of STFC, part of UK Research and Innovation.

\section*{Data Availability}

The radio data discussed in this paper are available in Table \ref{tab:rad_obs}. The data that forms the \textit{Swift}-XRT light curve is available from the \textit{Swift} Burst Analyser.



\bibliographystyle{mnras}
\bibliography{example} 






\bsp	
\label{lastpage}
\end{document}